\documentclass{aa}  
\usepackage{natbib}
\usepackage{xcolor}
\usepackage{graphicx}
\usepackage{natbib}
\bibpunct{(}{)}{;}{a}{}{,}
\usepackage{txfonts}
\newcommand{\blos}{B$_{\text{LOS}}$}
\newcommand{\rmref}{RM$_{\text{ref}}$}
\newcommand{\rmon}{RM$_{\text{ON}}$}
\newcommand{\rmoff}{RM$_{\text{OFF}}$}
\newcommand{\Av}{A$_{\text{V}}$}

\usepackage{makecell}

\begin{document}

\title{Helical Magnetic Fields in Molecular Clouds?}
 \subtitle{A New Method to Determine the Line-of-Sight Magnetic Field Structure in Molecular Clouds}
   \author{M. Tahani\inst{1},
R. Plume\inst{1},
J. C. Brown
          \inst{1}
          \and
          J. Kainulainen \inst{2,3} 
          }
      
   \institute{Physics \& Astronomy, University of Calgary, Calgary, Alberta, Canada\\
              \email{mtahani@ucalgary.ca}
	\and Dept. of Space, Earth and Environment, Chalmers University of Technology, Onsala Space Observatory, 439 92 Onsala, Sweden
    \and Max-Planck-Institute for Astronomy, Königstuhl 17, 69117 Heidelberg, Germany}
	
   \date{Received ; accepted }
   
\titlerunning{Helical Magnetic Fields in Molecular Clouds?}
\authorrunning{M. Tahani et al.}
            
\abstract
{Magnetic fields pervade in the interstellar medium (ISM) and are believed to be important in the process of star formation, yet probing magnetic fields in star formation regions is challenging.} {We propose a new method to use Faraday rotation measurements in small scale star forming regions to find the direction and magnitude of the component of magnetic field along the line-of-sight. We test the proposed method in four relatively nearby regions of Orion A, Orion B, Perseus, and California.} {We use rotation measure data from the literature. We adopt a simple approach based on relative measurements to estimate the rotation measure due to the molecular clouds over the Galactic contribution. We then use a chemical evolution code along with extinction maps of each cloud to find the electron column density of the molecular cloud at the position of each rotation measure data point. Combining the rotation measures produced by the molecular clouds and the electron column density, we calculate the line-of-sight magnetic field strength and direction.} {In California and Orion A, we find clear evidence that the magnetic fields at one side of these filamentary structures are pointing towards us and are pointing away from us at the other side. Even though the magnetic fields in Perseus might seem to suggest the same behavior, not enough data points are available to draw such conclusions. In Orion B, as well, there are not enough data points available to detect such behavior. This behavior is consistent with a helical magnetic field morphology. In the vicinity of available Zeeman measurements in OMC-1, OMC-B, and the dark cloud Barnard 1, we find magnetic field values of $-23\pm38~\mu$G, $-129\pm28~\mu$G, and $32\pm101~\mu$G, respectively, which are in agreement with the Zeeman Measurements.\ }{}{}

 
\maketitle

\section{Introduction}

While the exact role of magnetic fields in star formation is not clearly understood, they are known to be ubiquitous in the ISM and star forming regions. Many correlations between magnetic fields and star forming regions or filamentary structures have been observed in a variety of different surveys \citep[e.g.][]{PlanckXXXV, Goldsmith2008}. One of the proposed morphologies is that of helical magnetic fields threading molecular clouds/filaments \citep[e.g.][]{PressRelease, Heiles1987, ShibataMatsumoto1991, Nakamuraetal1993, Hanawaetal1993, Matsumoto1994, JohnstonBally1999, Hoqetal2017, Matthews2001, FiegePudritzI2000,FiegePudritzII2000, FiegePudritz2000APJ, Contrerasetal2013, StutzGould2016, SchleicherStutz2017}, an idea which, so far, has lacked systematic observational confirmation.

Observations of magnetic fields in Molecular Clouds (MC) have been made using the dust alignment method \citep{Anderssonetal2015, Palmeirim2013, Goldsmith2008, PlanckXXXV}, and Zeeman measurements \citep{Crutcher1999APJ520, Crutcher2005, Lietal2015, Kirketal2015}. The dust alignment method only provides a component of the magnetic field, which is projected on the plane of sky (perpendicular to the line-of-sight). Thus, observations to obtain the line-of-sight component of magnetic fields (\textbf{B}$_{\text{LOS}}$) are necessary to obtain information about the 3D structure of magnetic fields in MCs. 

Zeeman measurements do provide \textbf{B}$_{\text{LOS}}$, however, there are not enough Zeeman observations of MC available. As a consequence of relatively weak magnetic fields seen in MCs (e.g., 10s of $\mu$G), the splitting happens with very small frequency variations between the right and left circularly polarized components \citep{TrolandHeiles1982}, and even with very high signal to noise ratios, the frequency difference might still get masked \citep{Killeeetal1992}. They are also very time consuming \citep{CrutcherTroland2008}, and specifically require long telescope integration time in regions with relatively small magnetic fields \citep{RobishawThesis}.

The observed \citep[e.g.][]{Lietal2014, Palmeirim2013, Goldsmith2008} and theoretical \citep[e.g.][]{Pudritzetal2014, VanLoo2014, Khesalietal2014, Klessenetal2017} links between magnetism and star formation, coupled with the observational difficulties in measuring the magnetic field, point to a need for a new technique of detecting magnetism in star forming regions. 

We propose a new method to find \textbf{B}$_{\text{LOS}}$ in MCs based on Faraday rotation measurements. We test our method in four relatively nearby MCs: the Orion Molecular Cloud A (OMC-A ; the entire southern complex), the Orion Molecular Cloud B (OMC-B ; the entire northern complex), the Perseus Molecular Cloud (PMC), and the California Molecular Cloud (CMC) and find good agreement with available Zeeman measurements. We find that the magnetic field morphology in Orion A and California are consistent with a helical (or toroidal) magnetic field.

\subsection{Faraday Rotation}
When propagating through a magnetized region with free electrons, the plane of polarization of a linearly polarized electromagnetic wave will undergo Faraday rotation of $\Psi$ [rad], given by 
\begin{equation}
\label{FaradayRot}
\Psi = \lambda^2 \bigg(0.812 \int n_{\text{e}} \mathbf{B \cdot dl} \bigg) = \lambda ^2 \rm \text{RM} \rm~[rad],
\end{equation}
where $\mathbf{B}$ [$\mu$G] is the magnetic field, $\lambda$ [m] is the wavelength of the electromagnetic wave, $\mathbf{dl}$ [pc] is the pathlength through the magnetized region, and $n_{\text{e}}$ [cm$^{-3}$] is the electron density of the region. The integral value in brackets defines a quantity known as the rotation measure \citep[RM; e.g. ][]{Brown2008}. 

Faraday rotation has been widely used to investigate the large-Galactic-Scale magnetic field \citep[e.g.][]{SimardKronberg1980, Hanetal2006, Brownetal2007, Sunetal2008, VanEcketal2011, Ordogetal2017}, and to study the magnetic field of diffuse low-extinction filaments \citep{StilHryhoriw2016}. 
Previous attempts to study magnetic fields in high-extinction MCs using Faraday rotation have been performed by a number of authors \citep{WollebenReichb2004, Reichetal2002, WollebenReich2004,YusefZadehetal1997}. For example, \cite{WollebenReichb2004} utilize the concept of a Faraday screen - an object that can change the polarization angle and intensity of the polarized background - to estimate the field strength within the region. However, their method relies on an imprecise estimate of the distance to the screen, uncertainty in the electron density, and likely an oversimplification of the shape of the screen itself.

To estimate magnetic fields in MCs, we use a slightly different approach that avoids these difficulties by using extinction maps to obtain the total column density, and a chemical evolution code to determine more reasonable estimates of the electron density within MCs. With this, we can then work backwards to determine what the magnetic field must be to create the observed Faraday rotation measurements. 

\subsection{Free Electrons in MCs}
Free electrons are necessary for Faraday rotation to occur. The Photodissociation Region (PDR) models \citep{HollenbachTielens1999, HollenbachTielens1995, HollenbachTielens1997} predict the existence of  free electrons even in dense regions of MCs, and observations support the existence of free electrons in these regions \citep{Harrisonetal2013, Floweretal2007}. 

Most of the ISM is not illuminated by strong UV fields and this fact led to the belief that, in high column density regions in typical MCs, the UV field is so strongly attenuated that free electrons should be rare. Therefore Faraday rotation was not expected to occur within MCs. 

Cosmic Rays (CR), however, are known to be an important source of ionization in both diffuse and dense MCs \citep{Berginetal1999, Williamsetal1998, PadovaniGalli2013, Padovanietal2009, EverettZwibel2011, MorlinoGabici2015, Morlinoetal2015, BerginLanger1995, WillacyWilliams1993, 1993HasegawaHerbst} and thus CR ionization is an important source of producing free electrons in MCs. Calculating the CR ionizing factor, $\zeta$, is not straight forward and this factor may not be linear throughout the entire cloud \citep{Padovanietal2016, Morlinoetal2015, PadovaniGalli2013}. However, for the resolution and the scales that we are interested in, we assume it is constant\footnote{See \cite{EverettZwibel2011} for a history of theoretical studies of CR penetration into MCs.}.  With the confirmed existence of free electrons in MCs, we can expect that Faraday rotation occurs in MCs, as well as in the rest of the ISM. 

\section{Data}
\label{data}
Our method uses RMs of extragalactic sources with lines-of-sight passing near and through individual MCs to extract the strength and direction of magnetic fields in environments local to these MCs. Below we describe the RM data and the extinction maps that we use in our method. 

\subsection{Rotation Measure Catalog}
We use the RM values from \citet[][hereafter TSS09]{Tayloretal2009} catalog. They obtain RMs for 37543 polarized radio sources by reanalyzing the NRAO VLA Sky Survey data \citep[NVSS;][]{Condon1998}. For the regions of interest to us, within our specifically defined boundaries, TSS09 has 50 RMs within the OMC-A, 16 in OMC-B, 35 in PMC, and 43 in CMC. Fig.~\ref{AllTaurusRM} shows the map of RM data points in the PMC and CMC, and Fig.~\ref{AllOrionRM} shows the map of RM data points in OMC-A and OMC-B. The diameter of the RM circles is proportional to the magnitude of the RM; blue (red) circles indicate positive (negative) RMs, where the average line-of-sight magnetic field is directed towards (away from) us. The background color image represents the visual extinction map (see Sec. \ref{ExtinctionData}), with brighter or green color showing greater extinction.

\subsection{Extinction Map}
\label{ExtinctionData}
To map the hydrogen (HI + H$_2$) column density of each MC, we use visual extinction maps (in units of magnitudes of visual extinction or A$_{\text{V}}$) provided by \citet[][hereafter KBHP09]{Kainulainenetal2009}. They obtained near-infrared dust extinction maps using the 2MASS data archive and the NICEST \citep{Lombardi2009Nicest} color excess mapping technique. These maps have been produced with an arbitrary  physical resolution of 0.1~pc, which is the Jeans length for a core at T = 15~K and mean particle density of $\overline{n} = 5 \times 10^4~\text{cm}^{-3}$. We use these extinction maps as a proxy for N$_{\text{HI+H}_2}$ (as well as for obtaining electron abundances), by applying the \cite{Bohlinetal1977} conversion factor.

\section{Methodology}
\label{method}

The RMs in TSS09 are the result of polarized radiation passing through the entire line-of-sight of the Galaxy, from the source to the receiver (on Earth). Since we wish to recover the component of the RM which is produced by only the MC, we need to decouple the Faraday rotation produced by the Galaxy from that produced within the MC itself.

To accomplish this, we divide the integral in equation \ref{FaradayRot} into two parts: the contribution from the MC (RM$_{\text{MC}}$) and the contribution from everything else along the line-of-sight (Galactic contribution, RM$_{\text{Gal}}$). The Galactic contribution to the RM can be estimated by using RMs from positions that fall near the MC but are  far enough away that they are clearly not affiliated with it.  We refer to these as OFF positions, and designate their rotation measures as RM$_{\text{OFF}}$. 

\begin{figure*}
\vspace{5cm}
\centering
\includegraphics[scale=0.4]{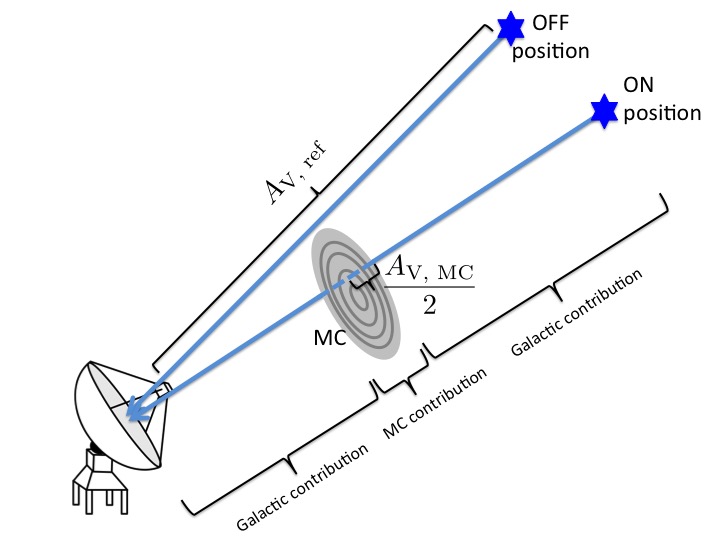}
\caption{Schematic illustrating ``ON'' and ``OFF'' RM positions, relative to a molecular cloud (MC). To find the magnetic field in the molecular cloud (MC) we need to disentangle the RM produced by the Galaxy from that produced by the MC itself. We do so by subtracting the rotation measure (RM) of a nearby point  called the OFF position, from the ON position which has an RM produced by both the MC (MC contribution) and the Galaxy (Galactic contribution). See Sec. \ref{refPoints} for details. Additionally, we need to consider the effects of all the layers of the MC from the exterior to the center of the cloud to reach to extinction value of $\text{A}_{\text{V, MC}} = \text{A}_{\text{V}} - \text{A}_{\text{V, ref}}$. However, since the cloud is symmetrical along the line-of-sight, and is illuminated from both sides by an ambient UV field, we assume the center of the cloud has an extinction of  $\frac{A_{\text{V, MC}}}{2}$ magnitudes.}
\label{PathlengthLayer}
\end{figure*}

\begin{figure*}
\vspace{3cm}
\centering
\includegraphics[scale=0.7, trim={2cm 6.5cm 4.1cm 1.8cm},clip]{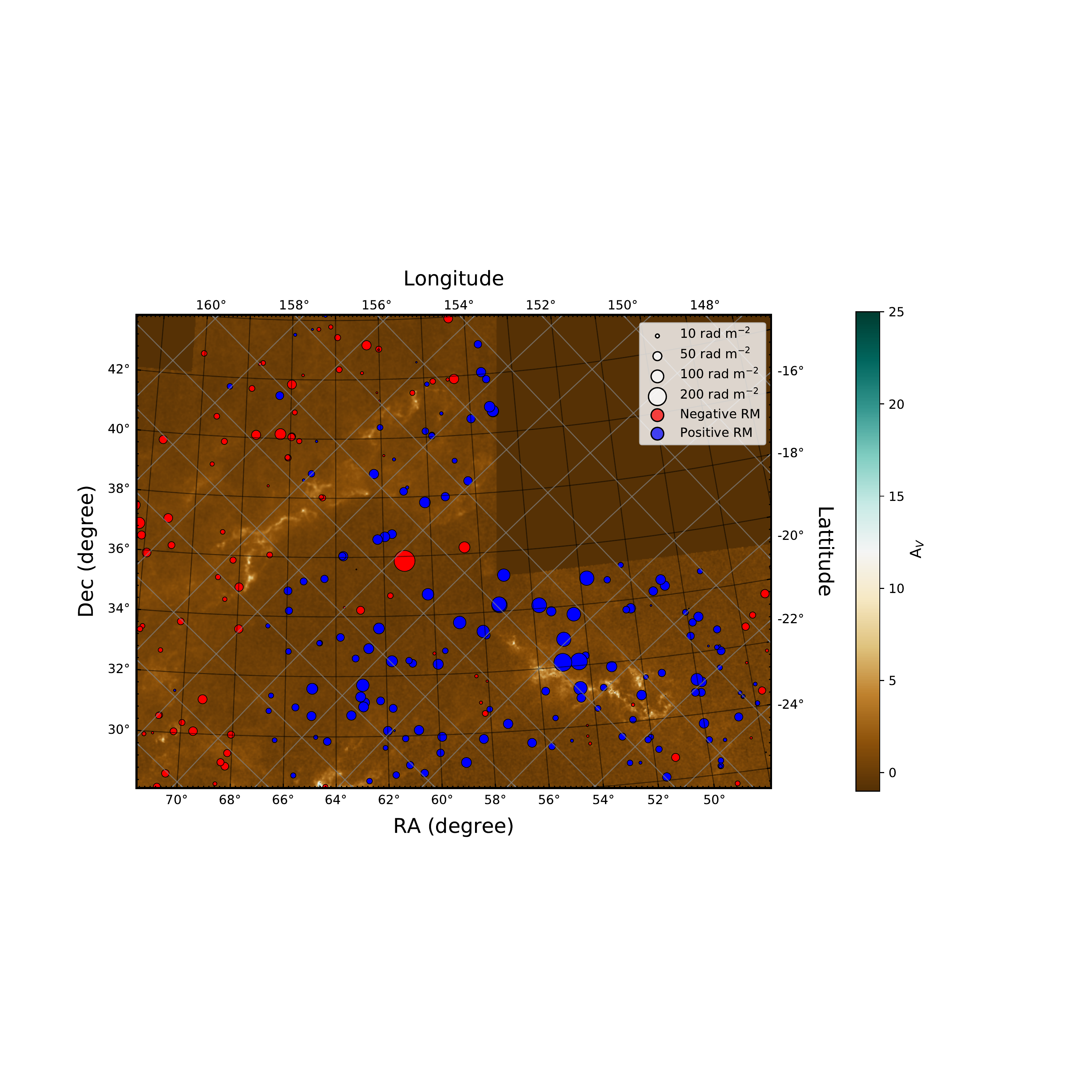}
\caption{Rotation measure (RM) values from the catalog of TSS09 mapped on the extinction map of CMC and PMC. Blue (red) circles indicate positive (negative) RM values. The size of the circles is proportional to the magnitude of the RM. The gray grid provides galactic coordinates, whereas the black grid provides equatorial coordinates. Color image shows the extinction map (A$_{\text{V}}$) in units of magnitudes of visual extinction provided by KBHP09.}
\label{AllTaurusRM}
\end{figure*}

\begin{figure*}
\centering
\includegraphics[scale=0.7, trim={4cm 0.3cm 4cm 0cm},clip]{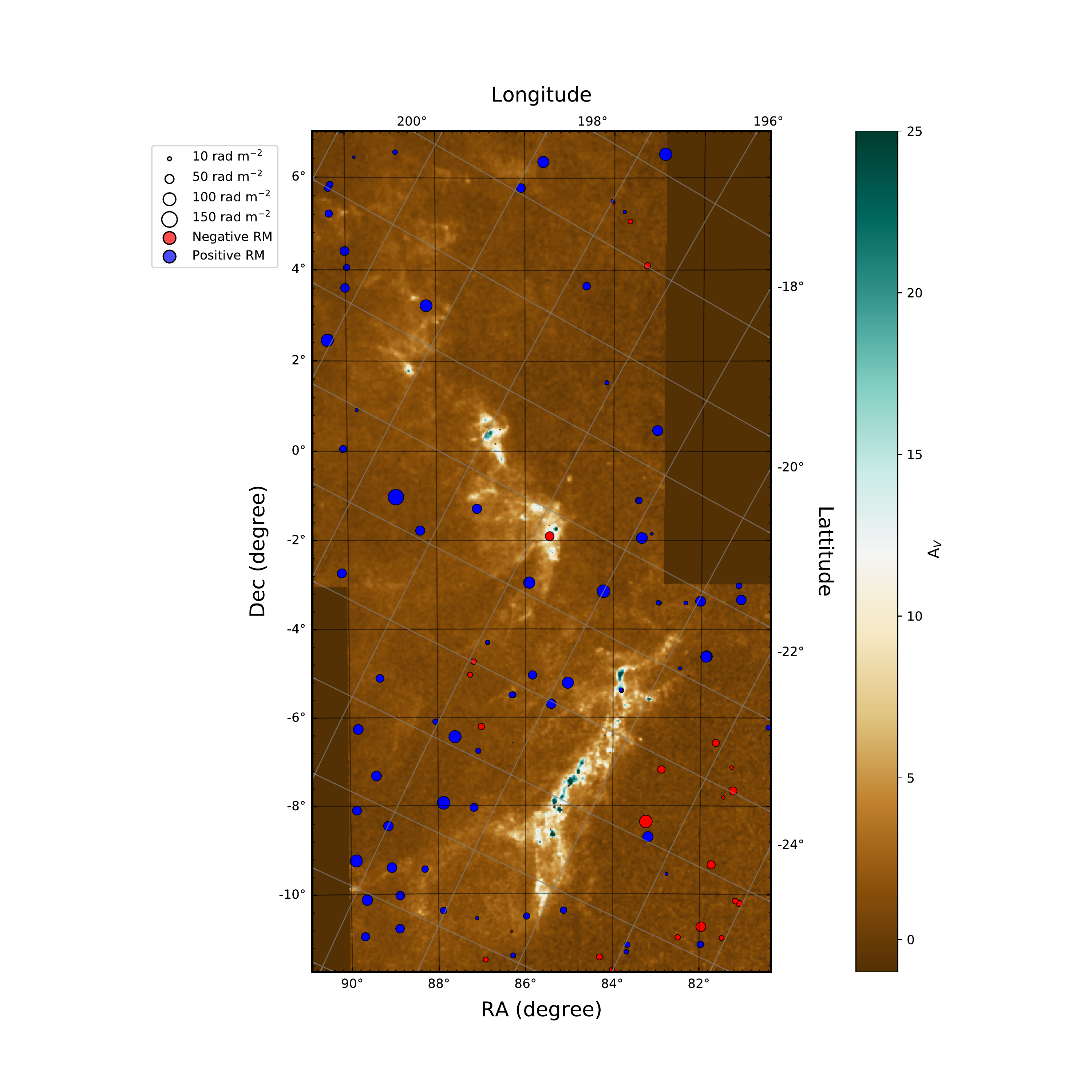}
\caption{Rotation measure (RM) values from the catalog of TSS09 mapped on the extinction map of Orion. OMC-A is the complex to the south and OMC-B is the complex to the north. Blue (red) circles indicate positive (negative) RM values. The size of the circles is proportional to the magnitude of the RM. The grey grid provides galactic coordinates, whereas the black grid provides equatorial coordinates. Color image shows the extinction map (A$_{\text{V}}$) in units of magnitudes of visual extinction provided by KBHP09.}
\label{AllOrionRM}
\end{figure*}

RM$_{\text{ON}}$ refers to any rotation measure in the TSS09 catalog that lies directly on or very near the MC (see Fig.~\ref{PathlengthLayer}). Since the angular separation between any RM$_{\text{OFF}}$ and RM$_{\text{ON}}$ is small compared to the angular size of the Galaxy, we assume that RM$_{\text{ON}}$ and RM$_{\text{OFF}}$ are essentially sampling the same pathlength through the Galaxy. 

We can then write RM$_{\text{ON}}$ as:
\begin{equation}
\text{RM}_{\text{ON}} = \text{RM}_{\text{MC}} + \text{RM}_{\text{Gal}}.
\end{equation}
Comparing RM$_{\text{OFF}}$ with RM$_{\text{Gal}}$ in Fig.~\ref{PathlengthLayer} shows that the pathlength of RM$_{\text{OFF}}$ is larger than that of RM$_{\text{Gal}}$ by a value equal to the pathlength through the cloud (i.e. a cloud-sized patch of the ISM, which we denote as RM$_\text{cloud-sized~ISM}$). In theory, we should account for this by subtracting the effects of this patch of the ISM from RM$_{\text{OFF}}$. We could do this by assuming that RM$_\text{cloud-sized~ISM}$ corresponds to a region with the same size as the MC but with the characteristics of the general ISM. However, We suggest that for dense clouds (MCs), RM$_{\text{cloud-sized ISM}}$ is negligible compared to RM$_{\text{MC}}$. To compare these two RMs, we examine the average values of $n_{\text{e}}$ and \textbf{B} of a typical MC with those of general ISM.

Average electron abundances for a typical MC, with density of around $\overline{n} =  10^3~\text{cm}^{-3}$, is roughly $10^{-4}$-$10^{-5}$ \citep{Harrisonetal2013}. The multiplication of these two yields  electron densities of $10^{-1}~\text{cm}^{-3}$-$10^{-2}~\text{cm}^{-3}$. The average density of the general ISM is $\overline{n} =  1~\text{cm}^{-3}$ \citep{MckeeOstriker2007}, with an average electron abundance of $\simeq 10^{-2}$ \citep{Cox2005}, which together provide an average electron density of $10^{-2}~\text{cm}^{-3}$. Thus, the average electron density of MCs can be 1 to 10 times that of the general ISM. 
The magnetic field strengths within MCs are, often, at least ten times higher than that of the general ISM \citep{PlanckXXXV}. Therefore, the contribution of the MC to the RM along the ``ON'' line-of-sight will be roughly 100 times larger than the ISM contribution of a similar size patch. Thus, for simplicity, we neglect the effect of the RM$_{\text{cloud-sized ISM}}$, and assume RM$_{\text{OFF}}$ is equal to RM$_{\text{Gal}}$. Note, however, that  this assumption may not be valid for all Galactic clouds and, in particular, in diffuse clouds RM$_\text{cloud-sized~ISM}$ may have to be specifically included.

We can, therefore, obtain the RM produced by just the MC by subtracting the OFF position from the ON position, i.e.:
\begin{equation}
\text{RM}_{\text{ON}} - \text{RM}_{\text{OFF}} = \text{RM}_{\text{MC}} =  \bigg( 0.812 \int n_{\text{e}} B_{\text{LOS}}dl \bigg)_{\text{MC}},
\end{equation}
where we have replaced the dot product by using the line-of-sight (LOS) component of the magnetic field within the MC (B$_{\text{LOS}}$). Furthermore, if we assume the magnetic field through the MC is uniform, we can extract B$_{\text{LOS}}$ from the integral
\begin{equation}
\text{RM}_{\text{MC}} =   \bigg( 0.812 \text{B}_{\text{LOS}} \int n_{\text{e}} dl  \bigg)_{\text{MC}}.
\end{equation}

Since $\int n_{\text{e}} dl $ is the electron column density in the MC, N$_{\text{e}}$, we can then write:
\begin{equation}
\text{RM}_{\text{MC}} =  \bigg( 0.812 \text{B}_{\text{LOS}} ~\text{N}_{\text{e}}  \bigg)_{\text{MC}}.
\label{FaradayFinal}
\end{equation}
Accordingly we can obtain:
\begin{equation}
\big(\text{B}_{\text{LOS}}\big)_{\text{MC}} = \frac{\text{RM}_{\text{MC}}}{0.812 \text{N}_{\text{e}} }.
\label{finalRMEq}
\end{equation}

To find B$_{\text{LOS}}$ in the MC we need to determine reasonable values for RM$_{\text{OFF}}$ and the electron column density (N$_{\text{e}}$), for each observed point.  This will be discussed below.

\subsection{Estimating RM$_{\text{OFF}}$ and RM$_{\text{ref}}$}
\label{refPoints}
We need to find suitable ON and OFF positions such that, when the RMs are subtracted, they isolate the effect of the MC alone. To find the ON positions, we search for RM measurements that visually fall on the MCs of interest (i.e. in higher column density regions). To find the OFF positions we hand pick a number of RMs that have low column densities (i.e. A$_{\text{V}} <1$) and are also far enough away from the cloud that they are clearly not directly related to it.  Therefore, in terms of A$_{\text{V}}$ and position, the OFF positions are associated with the general Galactic background rather than with the MC of interest.

Since variable Galactic structure can produce different RM values in different OFF positions, we use a number (N) of OFF positions to determine an ``average'' OFF position, which we call RM$_{\text{ref}}$, i.e. 
\begin{equation}
\text{RM}_{\text{ref}} = \sum_{i=1}^N \frac{\text{RM}_{\text{OFF, i}}}{N}.
\end{equation}

While we could use all of the positions in our maps that obey the above criteria
to produce RM$_{\text{ref}}$, we also wish to examine the magnetic field in the lower column density gas that immediately surrounds the MCs.
Therefore, we develop a method to determine the optimal number of OFF positions to incorporate into RM$_{\text{ref}}$.  This ensures that we have a robust and useful value for RM$_{\text{ref}}$ as well as leaving us enough RMs at the lower column density cloud edges to incorporate into our B field analysis. 

\begin{figure*}
\centering
\includegraphics[scale=0.7, trim={1cm 0cm 2cm 0cm},clip]{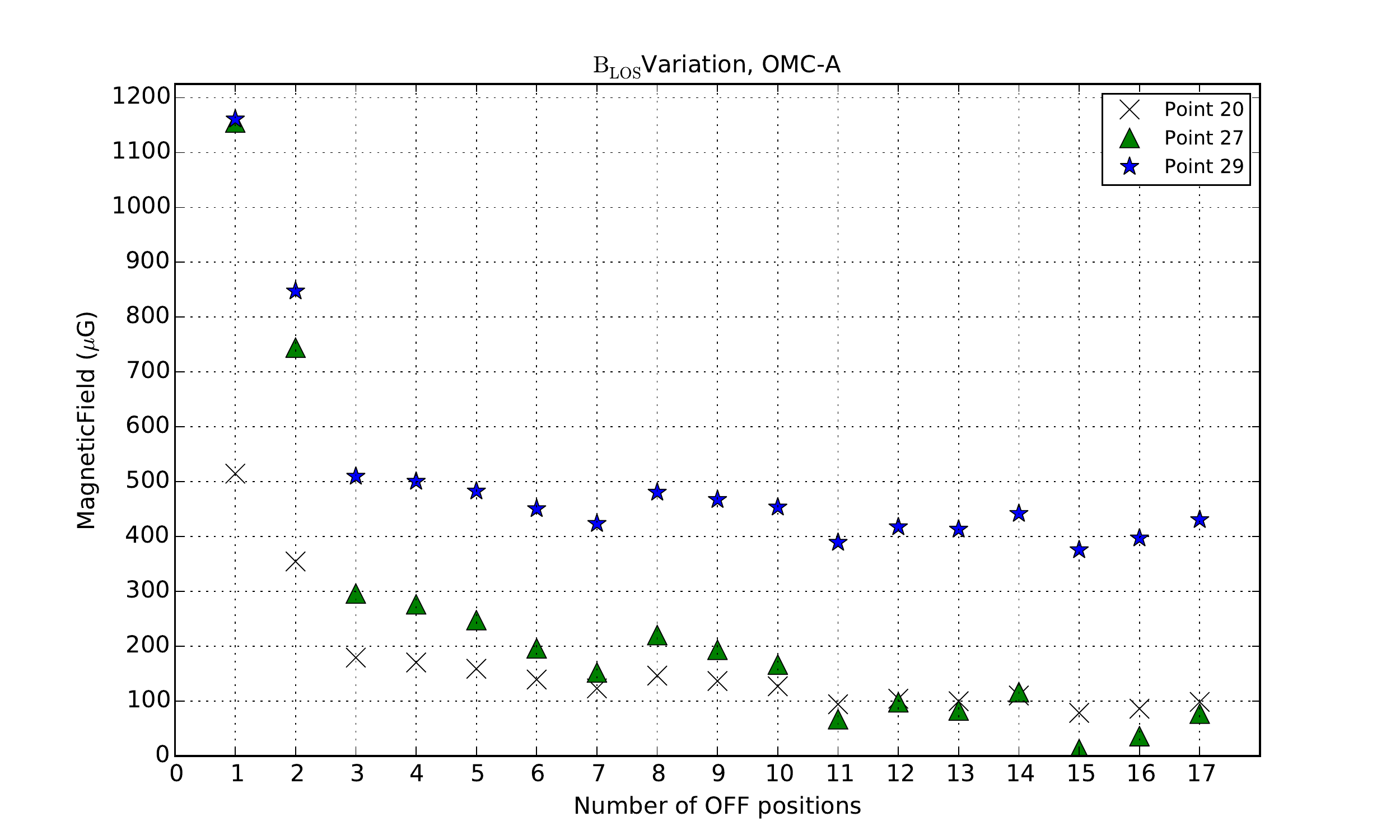}
\caption{Magnetic field values in OMC-A using different numbers of OFF positions to calculate the ``reference'' value (\rmref ).  Results are presented for three different points in Orion A. The x-axis indicates the number of OFF positions used in RM$_{\text{ref}}$. The y-axis shows the calculated magnetic field value. As discussed in the text, the magnetic field stabilizes at roughly 12 OFF positions.}
\label{BVariNRef}
\end{figure*}

For this purpose, we investigate how the derived magnetic field strength and direction changes as we increase the number of OFF positions from 1 to N. 
We find that, with few OFF positions, there is a large variance in the strength and direction of the derived magnetic fields. However, as we continue to increase the number of OFF positions, the variations decrease and the B field strengths and directions stabilize to a constant value (see Fig.~\ref{BVariNRef}). We choose the optimal number of OFF positions as the point at which the variance is minimized.  

From this analysis, we find the optimal number of reference points to be 12 for OMC-A, 5 for OMC-B, 8 for CMC, and 11 for PMC. The resultant values of RM$_{\text{ref}}$ are 1.4~rad~m$^{-2}$ for OMC-A, 32.3~rad~m$^{-2}$ for OMC-B, 4.0~rad~m$^{-2}$ for CMC, and 31.1~rad~m$^{-2}$ for PMC. Using the ``reference'' positions in lieu of a single OFF position, equation \ref{finalRMEq} becomes:
\begin{equation}
\big(\text{B}_{\text{LOS}}\big)_{\text{MC}} = \frac{\text{RM}_{\text{ON}}-\text{RM}_{\text{ref}}}{0.812 \text{N}_{\text{e}} }.
\label{finalRMEq2}
\end{equation}

\subsection{Obtaining the Electron Column Densities}
Determining the electron column density (N$_\text{e}$) in MCs requires an assessment of the total column density (N(HI + H$_2$)) through the MC, as well as a determination of the electron abundance as a fraction of total column density (X$_\text{e}$).  The former can be estimated from the KBHP09 extinction maps and the latter from a chemical model.  These steps will be outlined below.

\subsubsection{Determining Total Column Density from the Extinction Maps}
\label{ExtinctionFinding}
The KBHP09 maps are created by interpolating the measured extinction values onto a regular grid with a physical spacing of 0.1~pc between points. This corresponds to a different angular separation between points for different clouds due to their distances.  The RM measurements are not always at one of the precise positions where A$_{\text{V}}$ is tabulated.  Therefore, for each RM$_{\text{ON}}$ position we find the closest tabulated extinction point (within a distance of 0.1~pc) and assign it to that RM$_{\text{ON}}$.  This provides the extinction along the entire pathlength at each ON position (A$_\text{{V, ON}}$).

Since each chosen OFF position also has an associated A$_{\text{V}}$, 
after finding the optimum number of OFF positions and the value of RM$_{\text{ref}}$ for each MC, we also find the average extinction value of the reference position using:

\begin{equation}
\text{A}_{\text{V, ref}} = \sum_{i=1}^N \frac{\text{A}_\text{{V~OFF, i}}}{\text{N}}.
\end{equation}
The values of A$_{\text{V, ref}}$ are 0.45, 0.46, 0.54, and 0.67 for OMC-A, OMC-B, CMC, and PMC, respectively. To find the extinction value of the MC itself, we once again subtract the extinction in OFF position from that in the ON position.  i.e. :
\begin{equation}
A_{\text{V, MC}} = A_{\text{V, ON}} - A_{\text{V, ref}}. 
\end{equation}

\subsubsection{Determining the Electron Column Density}
In order to estimate the electron abundance, one requires a chemical model, which incorporates a number of relevant chemical reactions for particular gas conditions (density, temperature, UV field strength, cosmic ray ionization rate, etc.) and finds the abundance of each species as a function of time and depth (or A$_{\text{V}}$) within the cloud. We use an in-house chemical evolution code \citep[see][]{Gibsonetal2009}, which has been rigorously tested against the results of other established codes.

We utilize the UMIST Rate 99 database to obtain the reaction rates \citep{LeTeuffetal1999}. We use a small network of 229 gas-phase reactions coupling 28 different species including C, C$^+$, CO, O, O$_2$, CH, CH$^+$, CO$^+$, H, H$_2$, H$_2$O$^+$,  H$_3$O $^+$, HCO$^+$, O$^+$, etc. Additionally we include a simple treatment for gas-grain interactions via adsorption, thermal evaporation, CR desorption and photodesorption in manner outlined by \cite{BerginLanger1995} and \cite{1993HasegawaHerbst}. It does not include any surface-grain reactions.

Our chemical code assumes that each MC has a constant density and temperature, and is illuminated externally by a constant UV field (parametrized by G$_{\text{o}}$, where G$_{\text{o}}$ = 1 is the strength of the average ISM radiation field) and CR ionization rate.  For each of our clouds, we obtain the first three parameters from the literature. We assume  a constant CR ionization rate of $1.3 \times 10^{-17} \text{s}^{-1}$ for all clouds. 

The chemical code takes each cloud to be homogeneous and planar in structure, and is sliced into 100 layers of equal width. Each layer corresponds to a different depth into the cloud and therefore, we can calculate the amount of visual extinction (A$_{\text{V}}$) from the exterior (surface) of the cloud to the center of each layer. This controls how the external UV is attenuated as a function of depth, which, in turn, affects the importance of photo-reactions. In each layer, we start with standard cosmic abundances of all species and run our code until we achieve chemical equilibrium. The final outcome is a list of the equilibrium abundance of each species (including free electrons) as a function of extinction (depth) into the cloud. 

Even though we use a simple homogeneous, plane-parallel chemical model rather than a more sophisticated hydrodynamic approach \citep[e.g.][]{Seifried2017, ClarkGlover2014, Smithetal2014, GloverClark2012}, our simplified model provides similar electron abundances, when compared with these more detailed models.  For example, we find that our electron abundances are consistent with those of \cite{Gloveretal2010} in densities of 100~cm$^{-3}$ and 1000~cm$^{-3}$. This is true even though their initial temperature is different than ours. They, however, incorporate a cooling system that allows the high extinction parts of the gas to cool down to 10~K, which is close to the temperatures we have for our selected regions. Our electron abundances are also consistent with those from a variety of established PDR/chemical models (see the comparison by \cite{Rollingetal2007}). Thus, in this work, there does not seem to be much to gain from applying a more sophisticated approach to the chemical modeling.

Our chemical model calculates the electron abundance in each layer of the cloud.  However, to reach any given layer, we pass through all overlying layers which may have different electron abundances.
Consequently, to calculate the electron column density for a position in the MC with a given A$_{\text{V}}$, one cannot naively assume that the total electron column density is obtained directly from the electron abundance of one single layer multiplied by the total column density of that position (i.e, N$_{\text{e}} = \text{X}_{\text{e}} \times \text{N(HI+H}_2\text{)}$). We must, instead,  account for the contribution of each layer separately, since in each layer the electron abundance may be different due to the different UV attenuation. 

The total electron column density, N$_{\text{e}}$, is given by the equation:
\begin{equation}
\text{N}_{\text{e}} = \Sigma \text{N}_{\text{e, i}}, 
\end{equation}
where 
 N$_{\text{e, i}}$ is the electron column density in each layer and the sum is performed over all overlying layers from the surface of the cloud to the layer of interest.
 N$_{\text{e, i}}$, in turn is given by:
 \begin{equation}
\text{N}_{\text{e, i}} = X_{\text{e, i}} \times \text{N}_{\text{i}}(\text{HI + H}_2),
\label{abundToDens}
\end{equation}
where 
X$_{\text{e, i}}$ is the electron abundance in each layer, and is calculated by the chemical model.
 N$_{\text{i}}(\text{HI + H}_2)$ is the hydrogen column density in each overlaying layer. 
To evaluate N$_{\text{i}}(\text{HI + H}_2)$ in each layer we first subtract the extinction of that layer (A$_{\text{V, i}}$) from the extinction of the layer above (A$_{\text{V, i-1}}$) and then use the conversion factor of $2.21 \times 10^{21}$ by \cite{Bohlinetal1977} as follows:
\begin{equation}
\text{N}_{\text{i}}(\text{HI + H}_2) = (\text{A}_{\text{V, i}} - \text{A}_{\text{V, i-1}}) \times 2.21 \times 10^{21}.
\label{abundToDensLayers}
\end{equation}
Hence, the total electron column density along the line of sight from the surface of the MC to the layer of interest becomes:
\begin{equation}
\begin{aligned}
\text{N}_{\text{e}} = \Sigma \text{N}_{\text{e, i}} =   \Sigma \big( X_{\text{e, i}} \times (\text{A}_{\text{V, i}} - \text{A}_{\text{V, i - 1}})  \big) \times 2.21 \times 10^{21}.
\label{NeLayerTot}
\end{aligned}
\end{equation}

To assess which layers we must include in equation~\ref{NeLayerTot}, we presume the cloud is symmetrical along the line-of-sight as represented in Fig.~\ref{PathlengthLayer} and that the UV field is equally illuminating both sides of the cloud. However, the value of A$_{\text{V, MC}}$ obtained from the extinction maps is a measure of the extinction through the entire MC (front and back). Accordingly, in a MC with A$_{\text{V, MC}}$ = X, the total amount of UV attenuation from surface to center is only X/2. 

Our chemical model, however, assumes that the UV field is illuminating only one side of the cloud. Therefore, at a position where we measure A$_{\text{V, MC}}$ from the extinction maps, we only perform the sum in equation \ref{NeLayerTot} to a layer with  A$_{\text{V, MC}}$/2. Subsequently, we multiply the final sum by a factor of two to account for the fact that both the front and back sides of the MC contribute an equal amount to the total N$_{\text{e}}$ through the cloud. Thus, our final solution for N$_{\text{e}}$ is given by:
\begin{equation}
\label{extinctionLayer2}
\begin{aligned}
\text{N}_{\text{e}} = & 2 \times \Sigma \text{N}_{\text{e, i}} \\= & 2 \times \Sigma^{\frac{\text{A}_{\text{V, MC}}}{2}} \big((\text{A}_{\text{V, i}} - \text{A}_{\text{V, i - 1}}) \times X_{\text{e, i}} \big) \times 2.2 \times 10^{21}.
\end{aligned}
\end{equation}

Using Equations \ref{finalRMEq2}  and \ref{extinctionLayer2} along with the output from our chemical models and the measured ON and reference rotation measures therefore, enables us to calculate the line-of-sight magnetic field strength and direction in MCs.

The following example illustrates how this is done practically. Consider a particular point (for the purposes of this example, Point 22 in Fig.~\ref{OrionBottomAMap} in OMC-A, at $\alpha(J2000) = 86.31^{\circ}$, $\delta (J2000) = -5.49^{\circ}$). In the TSS09 catalog this position has an RM value of 23.5$\pm$9.5~rad~m$^{-2}$.  From the KBHP09 maps, this position has  an extinction value of A$_{\text{V}}$ = 2.84 mag. 

Using values for density, temperature, UV field strength, etc. found from the literature (see Sec. ~\ref{Orion} for details) we run our chemical code to generate a list of abundances  as a function of extinction (or depth) into the MC. 
Since A$_{\text{V, ref}}$ for OMC-A is 0.45, A$_{\text{V, MC}}$ for the MC at this point is  $2.84 - 0.45 = 2.39$.  Thus, in the output of the chemical code, we find the electron abundances in all layers from A$_{\text{V, MC}}$ = 0 to 1.20. The output of our code contains 5 layers to reach to this extinction value. These layers are (A$_{\text{V, i}}$, X$_{\text{e, i}}$) = (0.16, $1.40 \times 10^{-4}$), (0.46, $1.40 \times 10^{-4}$), (0.78, $1.41 \times 10^{-4}$), (1.08, $1.41 \times 10^{-4}$), and (1.40, $1.42 \times 10^{-4}$). The last layer, however, does not exactly match 1.20. Therefore, we interpolate between the last two layers, to find the electron abundance for a layer in between with an extinction value of 1.20. 

Subsequently, using equation~\ref{extinctionLayer2} we find N$_{\text{e}}$ as follows:
\begin{equation}
\begin{aligned}
\text{N}_{\text{e}} =  2 \times \bigg(& (0.16 - 0) \times 1.40 \times 10^{-4} \\&+ (0.46 - 0.16) \times 1.40 \times 10^{-4} \\ & + (0.78 - 0.46) \times 1.41 \times 10^{-4} \\& +  (1.08 - 0.78) \times 1.41 \times 10^{-4} \\ & +  (1.20 - 1.08) \times 1.41 \times 10^{-4} \bigg) \times 2.2 \times 10^{21} \\ =&  7.42 \times 10^{17}~\text{cm}^{-2}.
\end{aligned}
\end{equation}

Since the value of RM$_{\text{ref}}$ for OMC-A is 1.4~rad~m$^{-2}$ (Sec. \ref{refPoints}), $\text{RM}_{\text{ON}}$ - $\text{RM}_{\text{ref}}$ used in equation \ref{finalRMEq2} is 23.5 - 1.4 = 22.1 rad~m$^{-2}$. Finally, B$_{\text{LOS}}$ is calculated from equation~\ref{finalRMEq2} as:
\begin{equation}
\begin{aligned}
&\frac{22.1~\text{rad~m}^{-2}}{ 7.42 \times 10^{17}~\text{cm}^{-2} \times 3.24 \times10^{-19} \text{~pc~cm}^{-1}\times0.812}\\ =   &113~\mu G.
\end{aligned}
\end{equation}
Results of this calculation for all four MCs considered in this paper are provided in Tables \ref{OrionABValues}, \ref{OrionBBValues}, \ref{CaliforniaBValues}, and \ref{PerseusBValues}  and Figs. \ref{OrionBottomAMap} to \ref{PerseusBMap}.  Details and discussion for each MC are provided in Sec. \ref{result}.

\subsection{Uncertainty Analysis and Sensitivity Study}
\label{erroranalysis}
We carry out several analyses to determine how uncertainties in the chosen number of reference points, RM values, chemical code input parameters, positions, and extinction propagate into errors in the derived B$_{\text{LOS}}$. We discuss these below.

Since our \blos\ values are obtained using equation~\ref{finalRMEq2}, to estimate the uncertainties in our resultant \blos , we need to examine the uncertainties induced by both the cataloged RM values and the calculated N$_{\text{e}}$. 
Since the uncertainty in N$_{\text{e}}$ depends on our chemical model and its input parameters, we have to investigate the uncertainties that are caused by changing the input temperature and volume density. Additionally, since we positionally overlay the RM catalog and extinction maps, we have to account for any possible mismatches between the RM positions and the grid on which the extinction maps are produced.   This mismatch translates into a possible error in the value of A$_{\text{V}}$ assigned to any given RM point.  Therefore, our \blos\ values are a function of the cataloged RM values, A$_{\text{V}}$ (which is, itself, a function of the positional coordinates), and the chemical model input  density and temperature. Thus \blos\ is really parameterized by \blos\ (RM,~A$_{\text{V}}$,~n(HI+H$_2$), ~T), and has an uncertainty of:
\begin{equation}
\begin{aligned}
\delta \text{B}_{\text{LOS}} = \text{B}_{\text{LOS}} \bigg((\frac{\delta \text{RM}}{\text{RM}})^2  + (\frac{\delta \text{A}_{\text{V}}}{\text{A}_{\text{V}}})^2&& \\+ (\frac{\delta n({\text{HI}+\text{H}_2})}{n({\text{HI}+\text{H}_2})})^2 + (\frac{\delta \text{T}}{\text{T}})^2 \bigg)^{1/2}&&.
\label{beginUncertainty}
\end{aligned}
\end{equation}

\paragraph{Uncertainty In \blos\ from RM} 

In the RM catalog of TSS09, the source entries include sky position, Stokes I (total intensity), linear polarized intensity, fractional polarization, and RM, with estimated errors for each entry. Accordingly, each RM value in the TSS09 catalog has a corresponding RM uncertainty that we incorporate into our error bars. For the ON positions, we simply take the associated errors listed for those positions. The uncertainty in the RM$_{\text{ref}}$ measurement, however, is the standard deviation of the RM values of the chosen OFF positions. The \blos\ uncertainty from RM for each point is found as follows:
\begin{equation}
\begin{aligned}
\Delta \text{B}_{\text{RM}} = \text{B}_{\text{LOS}} \bigg(\frac{\delta (\text{RM}_{\text{ref}}) + \delta (\text{RM}_{\text{ON}})}{\text{RM}_{\text{ON}} - \text{RM}_{\text{ref}}}\bigg),
\end{aligned}
\end{equation}
where $\Delta \text{B}_{\text{RM}}$ is the uncertainty in \blos\ from RM, $\delta(\text{RM}_{\text{ref}})$ is the standard deviation of the OFF positions, $\delta(\text{RM}_{\text{ON}})$ is the tabulated uncertainty of the RM of the ON point. 

Considering OMC-A as an example once more, we calculate the standard deviation in RM$_{\text{ref}}$  for OMC-A to be 13.7~rad~m$^{-2}$ (using 12 OFF positions to calculate RM$_{\text{ref}}$).  Thus, RM$_{\text{ref}}$ in OMC-A is $1.4 \pm 13.7$~rad~m$^{-2}$.

Note that the uncertainties in the RMs tabulated in TSS09 and in the reference positions are the dominant source of error in our magnetic field calculation, and are the main reason that the uncertainties listed in Tables \ref{OrionABValues} to \ref{PerseusBValuesModif} are as large as they are. For instance, the tabulated value of RM$_{\text{ON}}$ of point 22 used in the example above is 23.5 $\pm$ 9.5~rad~m$^{-2}$, which results in large fractional errors in the derived magnetic field values. In TSS09 catalog there are also points like 21 (see Table ~\ref{OrionABValues}) with RM$_{\text{ON}}$ value of $-0.3 \pm 6.9$~rad~m$^{-2}$, which creates enormous relative uncertainty. This will be discussed further in Sec.~\ref{errordiscussion}.

\paragraph{Uncertainty In \blos\ From N$_{\text{e}}$}

Uncertainties in the electron column density are caused by uncertainties in our chemical code input parameters, since the density, temperature, and UV field strength may not be well characterized.

To investigate how changes in density affect the electron abundance, we hold all other input parameters constant and change the input volume density, (n(HI+H$_2$)), by $\pm 1 \%$, $\pm 2.5 \%$, $\pm 5 \%$, $\pm 10 \%$, $\pm 20 \%$, $\pm 30 \%$, $\pm 40 \%$, $\pm 50 \%$ from the cloud fiducial input density, n$_0$. We then rerun the chemical code with the altered density and obtain a new value for the electron abundance. Consequently, we obtain the value of B$_{\text{LOS}}$ for each point with the new electron abundances. We then calculate the B$_{\text{LOS}}$ differences from the original B$_{\text{LOS}}$ value. We denote these uncertainties in \blos~as $\Delta \text{B}_{n({\text{HI}+\text{H}_2})}$.

\begin{figure*}
\centering
\includegraphics[scale=0.48, trim={3.4cm 2.1cm 1.9cm 3.5cm},clip]{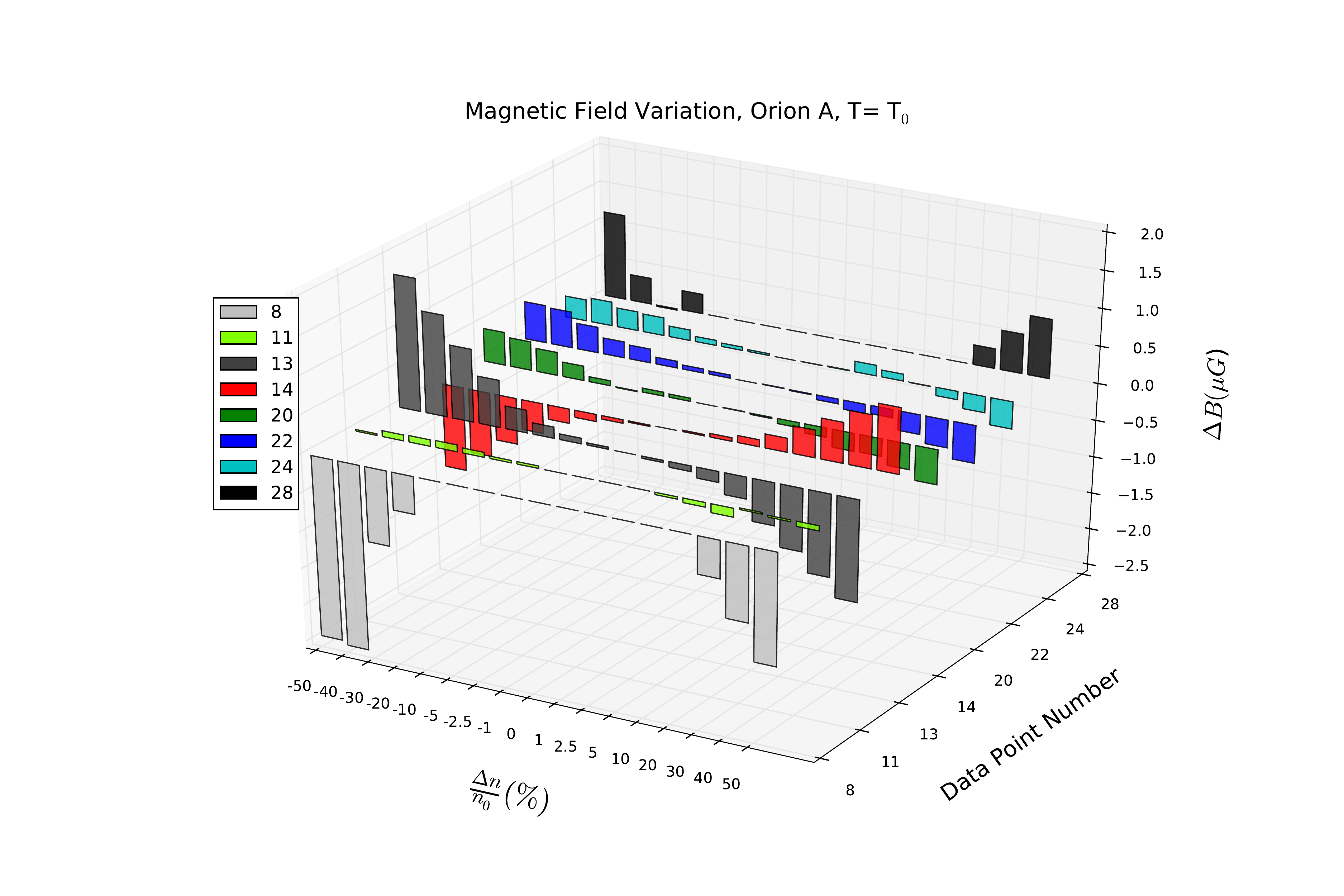}
\begin{tabular}{cc}
\includegraphics[scale=.48, trim={0.1cm 0.1cm 1cm 0.8cm},clip]{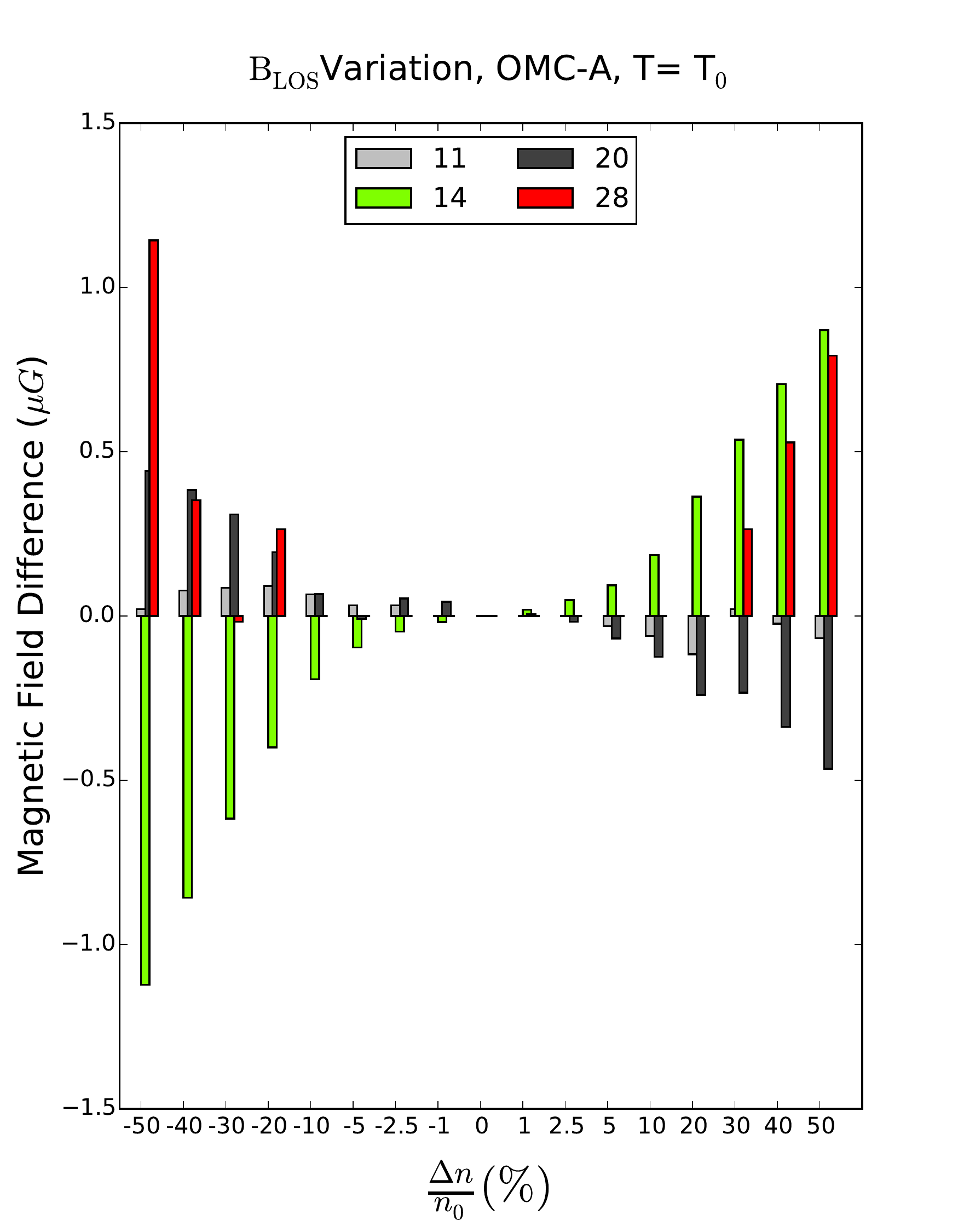}
\includegraphics[scale=.48, trim={0.1cm 0.1cm 1cm 0.8cm},clip]{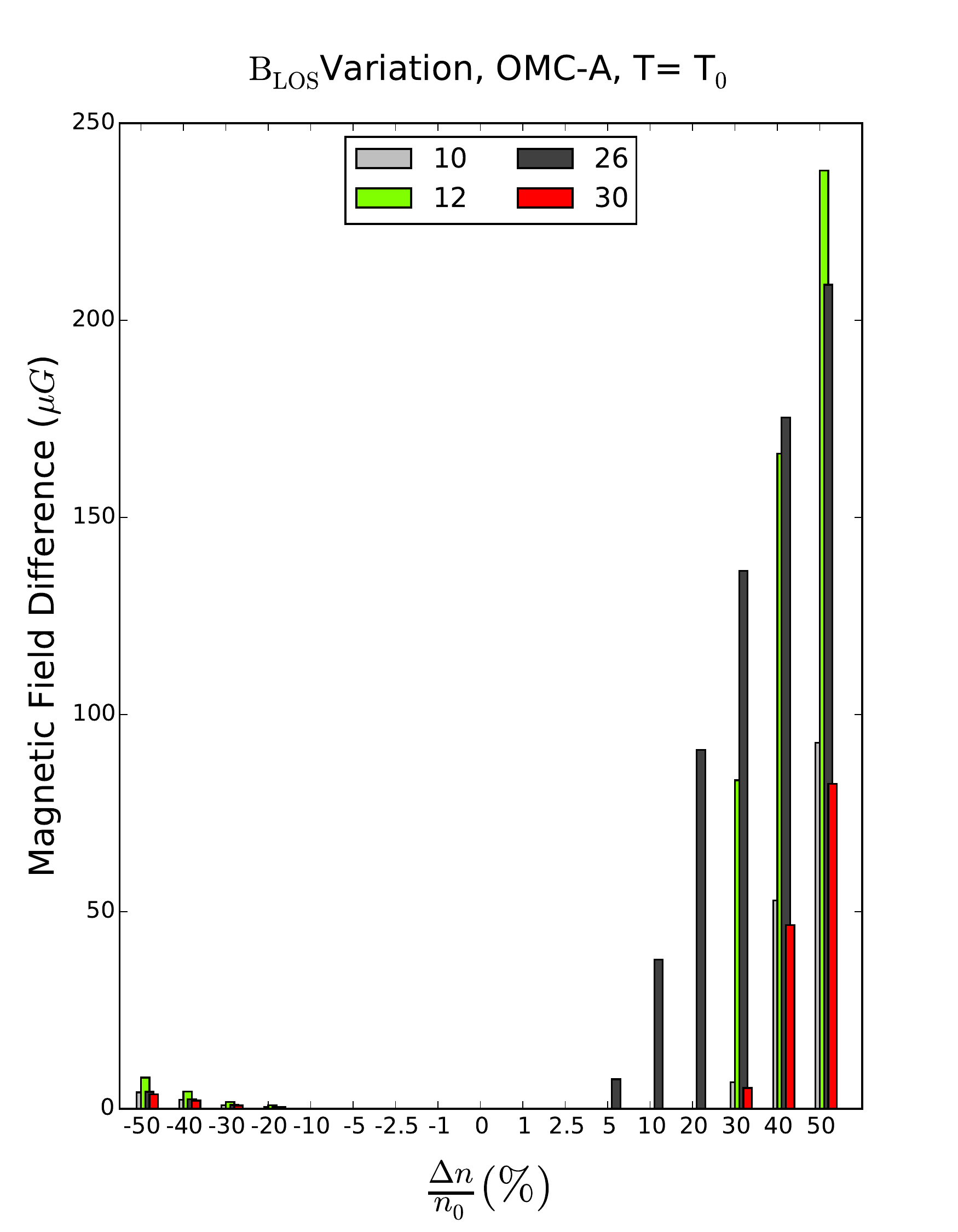}
\end{tabular}\\
\caption{Uncertainties in \blos\ due to uncertainties in the input volume density of the chemical evolution code. \textbf{Top Panel}: B$_{\text{LOS}}$ variation for a selection of positions in Orion A sampling regions with different A$_{\text{V}}$. The x-axis ($\frac{\Delta n}{n_0}$) indicates the relative ($\%$) changes in input density to the chemical model. The y-axis labels the data points presented and corresponds to the positions labeled in Fig.~\ref{OrionBottomAMap}. The z-axis plots the change in the magnetic field strength and direction from that calculated for the fiducial density $n_0 (10^4 cm^{-3})$. The \textbf{bottom left panel} shows \blos\ uncertainties for points with A$_{\text{V}}$ < 1. The \textbf{bottom right panel} shows uncertainties in \blos\ for a selection of points with A$_{\text{V}}$ >1.}
\label{OrionSensitivity}
\end{figure*}

\begin{figure*}
\centering
\includegraphics[scale=0.7, trim={8cm 0.3cm 0.8cm 0cm},clip]{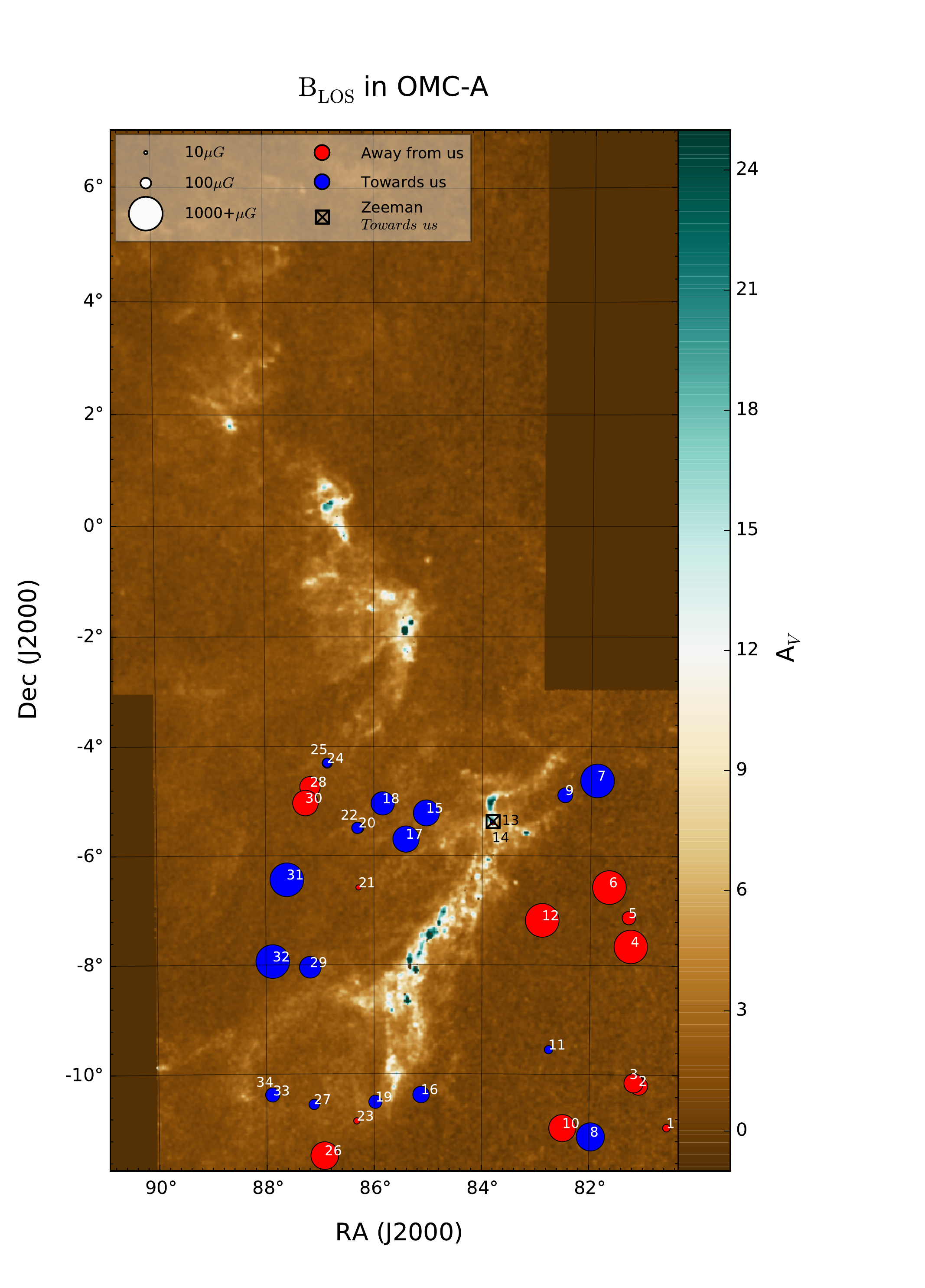}
\caption{\textbf{B}$_{\text{LOS}}$ in OMC-A. Blue (red) circles show magnetic fields toward us (away from us). The size of the circles indicate the magnitude of magnetic field. Black square shows the location of the available Zeeman measurements. Color image is the extinction map (A$_{\text{V}}$). The magnetic fields are dominantly towards us at the eastern side of this filamentary structure and away from us at its western side.}
\label{OrionBottomAMap}
\end{figure*}

\begin{figure*}
\centering
\includegraphics[scale=0.7, trim={8cm 0.3cm 0.8cm 0cm},clip]{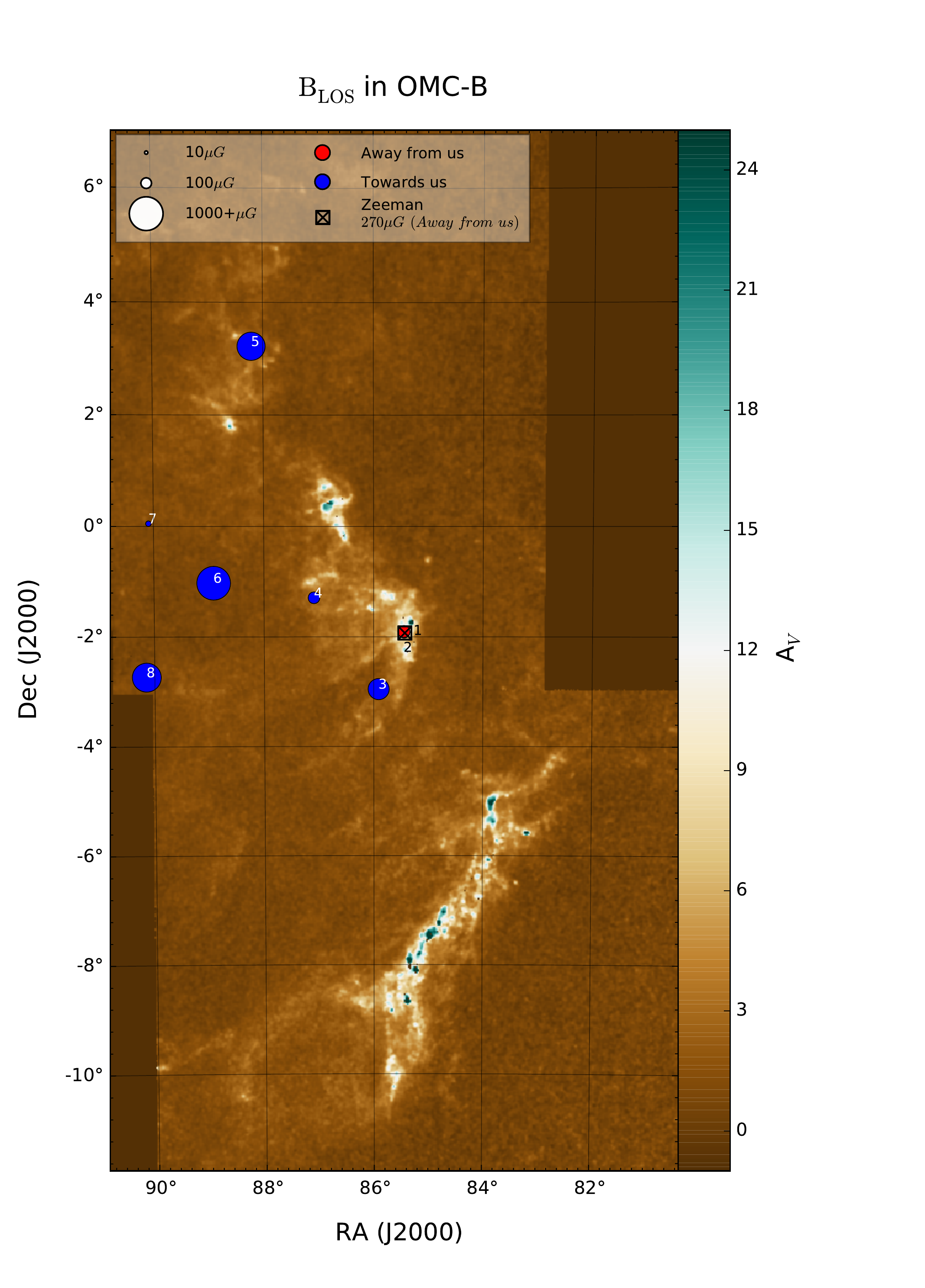}
\caption{\textbf{B}$_{\text{LOS}}$ in OMC-B. Blue (red) circles show magnetic fields toward us (away from us). The size of the circles indicate the magnitude of magnetic field. Black square shows the location of the available Zeeman measurements. Color image is the extinction map (A$_{\text{V}}$).}
\label{OrionTopBMap}
\end{figure*}

Referring back to OMC-A as an example, Fig.~\ref{OrionSensitivity} demonstrates how  B$_{\text{LOS}}$ changes  as the input density is varied. The top panel of Fig.~\ref{OrionSensitivity} shows B$_{\text{LOS}}$ deviations for a selection of data points in OMC-A. The z-axis indicates changes in  B$_{\text{LOS}}$ from the fiducial value (obtained from the fiducial input density n$_0$). The x-axis indicates the relative changes in the cloud initial (fiducial) density, and the y-axis indicates particular data points in OMC-A  as mapped in Fig.~\ref{OrionBottomAMap}. While we have performed this error analysis for every point, we only display a few select points for clarity.  The bottom left panel of Fig.~\ref{OrionSensitivity} represents variations in B$_{\text{LOS}}$ for data points with A$_{\text{V}} >1$, and the bottom right panel shows the same for data points with A$_{\text{V}} <1$. These figures show that B$_{\text{LOS}}$ variations are largest in the regions with lower visual extinction.  The main reason for this behavior is that, in low A$_{\text{V}}$ regions, the electron fraction is high and so changes in density result in relatively large  changes in N$_{\text{e}}$ which, in turn, affects B$_{\text{LOS}}$.  On the other hand, in the high A$_{\text{V}}$ regions, since we are looking through many cloud layers, changes in N$_{\text{e}}$ are averaged over many layers.

The resultant uncertainties in B$_{\text{LOS}}$ caused by changes in N$_{\text{e}}$  are asymmetrical and, therefore, we report magnetic field values in form of B$_{- \delta \text{B}} ^{+ \delta \text{B}}$, and in a case where the two  $\delta \text{B} $ are the same, in form of B $\pm \delta$B.

We carry out a similar analysis for the input temperature by varying it by  $\pm 5 \%$, $\pm 10 \%$, and $\pm 20 \%$ from the cloud fiducial input temperature, T$_0$, while holding the other parameters constant.  Similarly we obtain the electron abundance and therefore the new magnetic field values for each point. Changes to the input temperature introduce fairly small variations to B$_{\text{LOS}}$. We denote these uncertainties as $\Delta \text{B}_{\text{T}}$.

\paragraph{Uncertainty In \blos\ From Extinction and Position}

Since we have an uncertainty in matching the position between the RM catalog points and the grid on which the extinction maps are calculated (see Sec. \ref{ExtinctionFinding}), this translates into an error, $\Delta$B$_{\text{ext, coord}}$, in the assumed A$_{\text{V}}$. This arises because, while we take the A$_{\text{V}}$ value that lies closest to the RM position, there may be more than one value of A$_{\text{V}}$ in a 0.1~pc radius surrounding the RM point. To estimate the influence that this  has on our derived magnetic fields, we calculate B$_{\text{LOS}}$ for the maximum and minimum A$_{\text{V}}$ that falls within a 0.1~pc radius around each RM position. 

\paragraph{Total B$_{\text{LOS}}$ Uncertainty}
After finding the individual uncertainties, we can find the total uncertainty, using equation~\ref{beginUncertainty}, by \citep{UncertaintyBook}:
\begin{equation}
\begin{aligned}
\delta \text{B}_{\text{LOS}} = \bigg((\Delta \text{B}_{\text{RM}})^2  + (\Delta \text{B}_{\text{ext, coord}})^2&& \\+ (\Delta \text{B}_{n({\text{HI}+\text{H}_2}}))^2 + (\Delta \text{B}_{\text{T}})^2 \bigg)^{1/2}&&,
\end{aligned}
\end{equation}
where $\Delta \text{B}_{\text{RM}}$ is the error in B produced by the RM uncertainties for each data point in the TSS09 catalog along with the reference RM, $\Delta \text{B}_{\text{ext,coord}}$ is the error in B produced by the uncertainty in the assumed extinction value, $\Delta \text{B}_{n(\text{HI}+\text{H}_2)}$ is the error in B produced by the uncertainties in the chemical code input density, and $\Delta \text{B}_{\text{T}}$ is the same due to uncertainties in  the input temperature. We believe that we have been quite conservative in estimating the total B$_{\text{LOS}}$ error and, therefore, the true error may indeed be smaller than those quoted in Tables \ref{OrionABValues}, \ref{OrionBBValues}, \ref{CaliforniaBValues}, and \ref{PerseusBValues} .

\section{Results}
\label{result}

We used the method described above for each of the four MCs in our sample (OMC-A, OMC-B, PMC, and CMC). We compared the results to existing Zeeman measurements to verify the validity of the method. We discuss our results for each of these regions below.

\subsection{The Orion Molecular Cloud}
\label{Orion}
The OMC is a well-studied, active star forming region with relatively strong magnetic fields \citep{Crutcher1999APJ520, Crutcher2010}. Some prominent regions in OMC are the Orion Nebula Cluster, L1641, NGC2026, and NGC2024 with distances of 388 $\pm$ 5~pc, 428 $\pm$ 10~pc, 388 $\pm$ 10~pc, and roughly 420~pc, respectively \citep{Kounkeletal2017}. 

Orion A and B are the two distinct giant molecular clouds in the OMC complex. OMC-A is located at $80^\circ<\alpha (J2000)<88^\circ$ and $-12^\circ<\delta (J2000)<-4^\circ$. OMC-B is located at $84^\circ<\alpha (J2000)<95^\circ$ and $-4^\circ<\delta (J2000)<4^\circ$. 
 
For both regions, we use n(HI+H$_2$) = $10^4$~cm$^{-3}$ \citep{Castetsetal1990, Dutreyetal1993, JohnstoneBally1999a, JohnstoneBally1999b}, T = 25~K \citep{Mitchelletal2001, JohnstoneBally2006, Ballyetal1991, Castetsetal1990, Schneeetal2014, Buckleetal2012}, and a UV field strength of G$_{\text{o}} = 10^4$ (where G$_{\text{o}} = 1$ is the strength if the average interstellar UV field) as input to our chemical models.

Using the methodology described above, we calculate B$_{\text{LOS}}$ for all the available RM points in OMC-A and B.  These results are presented in Figs. \ref{OrionBottomAMap} and \ref{OrionTopBMap}, in which the size of each filled circle represents the strength of B$_{\text{LOS}}$ and the color represents the direction (blue towards the observer and red away).
Derived values of B$_{\text{LOS}}$ for Orion A \& B are provided in Tables \ref{OrionABValues} and \ref{OrionBBValues}. The reason for the large uncertainties was discussed in Sec. \ref{erroranalysis}, and will be explored in more detail in Sec. \ref{errordiscussion}.

To examine the veracity of our method, we compare our derived magnetic field strengths to those determined from other well-known methods, such as Zeeman measurements. For these two regions several Zeeman measurements are available \citep{Trolandetal1986, Trolandetal1989, Crutcheretal1999APJ515, Crutcher1999APJ520, Crutcher1999APJ514, Crutcheretal1996, Verschuur1996, Crutcher2010}, and are graphically represented on Figs. 6 and 7 as black squares.

Note that conventionally the negative sign represents magnetic field towards us in Zeeman measurements and away from us in RM studies. For consistency between discussions of RM and Zeeman measurements, we adopt the convention that  --B$_{\text{LOS}}$ indicates a magnetic field directed away from the observer and a +B$_{\text{LOS}}$ indicates a magnetic field toward the observer.

\subsubsection{Strength and Morphology of B$_{\text{LOS}}$in Orion A \& B}
\label{OrionA}
There are a number of Zeeman measurements in OMC-A, most of which fall in vicinity of a high extinction region with approximate coordinates of $\alpha(J2000) \simeq 83.81^{\circ}, \delta(J2000) \simeq -5.37 ^{\circ}$. The magnetic fields inferred from these different studies have wildly different values and, often, large error bars e.g. 
 $+360 \pm 80~\mu$G \citep{Falgaroneetal2008, Crutcher1999APJ520, Crutcher2010}, 
 $-79 \pm 99~\mu$G \citep{Crutcheretal1996}, $-40 \pm 240~\mu$G \citep{Crutcher1999APJ514, Crutcher2010}, $+190 \pm 90~\mu$G \citep{Crutcher1999APJ514}, and $-80 \pm 100~\mu$G \citep{Crutcher2010}. These studies suggest that  the magnetic field in this region (including error bars) might have any strength from $+440~\mu$G  to $-280~\mu$G.

In comparison, using our technique we have two data points in this area (sources 13 and 14 in Table \ref{OrionABValues} and Fig.~\ref{OrionBottomAMap}) with magnetic field values of $-23 \pm 38~\mu$G and $+15 \pm 36~\mu$G, respectively. Given the large error bars in both our technique and the Zeeman measurements, as well as the large dispersion in the Zeeman values, we find it promising that: a) our magnetic field strengths and directions fall within the envelope of those determined via Zeeman measurements, and b) that our error bars for these positions are, in fact, smaller than those for the Zeeman measurements.  Thus, we suggest that there is qualitative agreement between our results and those from Zeeman measurements. Having said that, comparing our results to those of Zeeman measurements must be done cautiously, since they are possibly looking at different regions within the MC (see Sec.~\ref{ZeemanComparison}).

Given that there are many more RM observations across the Galaxy than there are Zeeman measurements, our technique can also provide useful insight into the morphology of the line-of-sight magnetic field in MCs.  For example, Fig.~\ref{OrionBottomAMap} suggests that the magnetic field on the eastern side of OMC-A is predominantly positive (blue), whereas on the western side it is negative (red). This particular pattern has been previously observed \citep{Heiles1997}, and interpreted as helical magnetic fields \citep[e.g.][]{JohnstonBally1999, Hoqetal2017, Matthews2001}. We will discuss this possibility in more detail in Sec.~\ref{HelicalB}. 

The two available Zeeman measurements in OMC-B are in a high extinction area at $\alpha(J2000) \simeq 85.44^{\circ}, \delta(J2000) \simeq -1.93^{\circ}$, and have significantly different magnetic field strengths and error bars,  e.g.   $-270 \pm 330~\mu$G \citep{Crutcher1999APJ514} and $-87 \pm 5.5~\mu$G \citep{Crutcheretal1999APJ515}. Our measurements in this proximity are points 1 and 2 (see Table \ref{OrionBBValues} and Fig. \ref{OrionTopBMap}) with magnetic field values of $-119 \pm 25~\mu$G and $-129 \pm 28~\mu$G (i.e., both pointing away from us). 
 
As with OMC-A, there is general agreement in both the direction and strength of magnetic field between the two Zeeman measurements and our own results.  There are, however, fewer RM points in OMC-B with which to infer the large-scale morphology of the magnetic field.

\subsection{California and Perseus}

It is important to test our method in different environment conditions besides the well known region of Orion. Thus, we test our method in the PMC and CMC, which have lower density and ambient UV field strengths than Orion.

\subsubsection{Strength and Morphology of B$_{\text{LOS}}$ in  the California Molecular Cloud (CMC)}
\label{California}

The CMC occupies a region of roughly 58$^\circ$<$\alpha (J2000)$<70$^\circ$ and 34$^\circ$<$\delta (J2000)$<42$^\circ$ \citep{Lombardietal2010}. It is part of the Gould Belt and has modest star formation activity \citep{Harveyetal2013}. \cite{Ladaetal2009} report a distance of 450 $\pm$ 23~pc to the cloud. The cloud extends around 80~pc and has a mass of around 10$^5$ M$_{\odot}$.

\begin{figure*}
\centering
\includegraphics[scale=0.7, trim={1.1cm 3.2cm 4cm 1.8cm},clip]{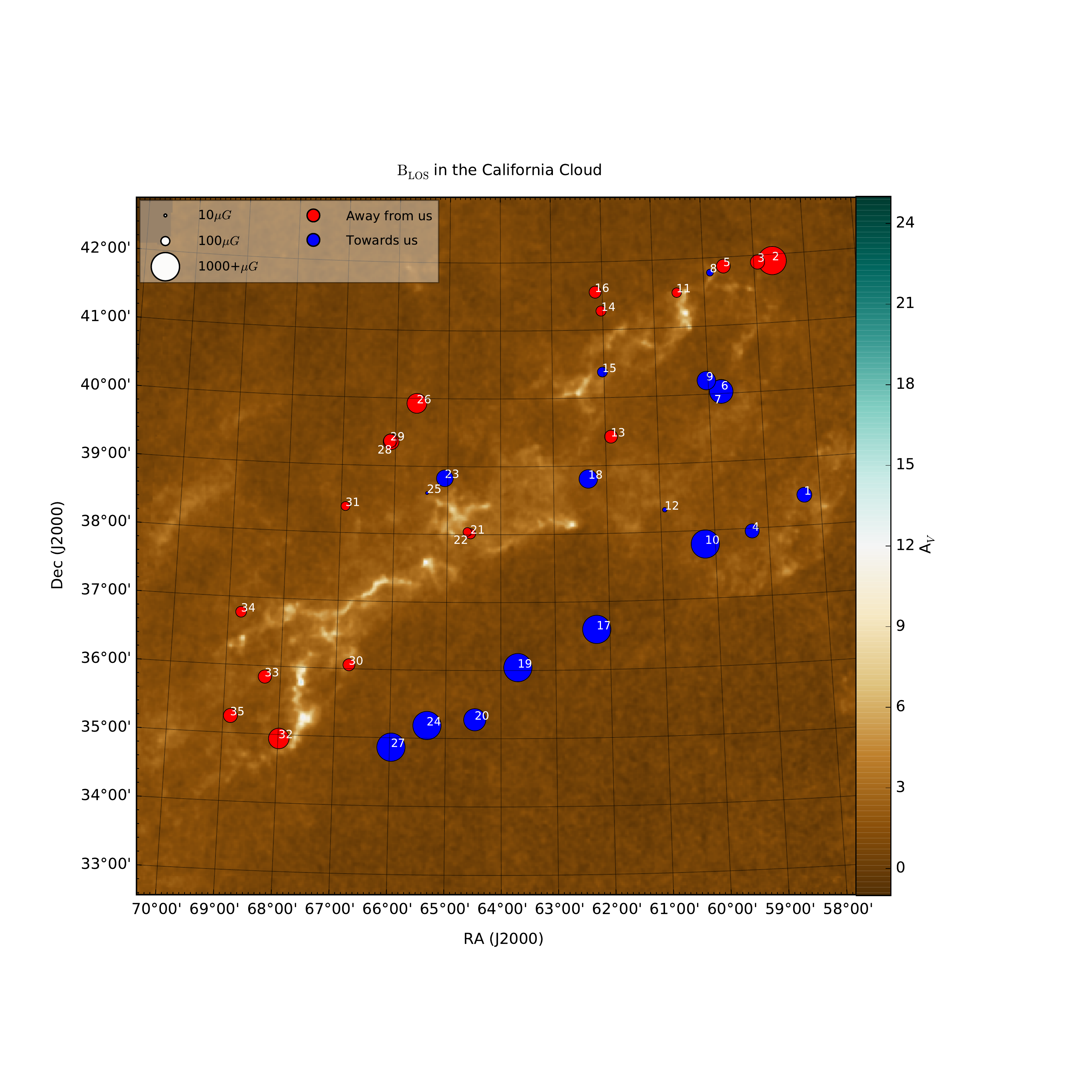}
\caption{\textbf{B}$_{\text{LOS}}$ in the California MC. Blue (red) circles show magnetic fields toward us (away from us). The size of the circles indicate the magnitude of magnetic field. Color image is the extinction map (A$_{\text{V}}$). The magnetic fields are dominantly towards us at the western side of this filamentary structure and away from us at its eastern side.}
\label{CaliforniaBMap}
\end{figure*}

Considering the results of \cite{Kongetal2015} and \cite{Ladaetal2009} we take an initial volume density of n(HI+H$_2$) = 450~$\text{cm}^{-3}$, a temperature of T = 10~K, and UV field radiation strength of G$_{\circ}$ = 1.0 for the input to our chemical models. Using the same method described in Sec. \ref{method}, we then calculate the magnetic field strength and direction in CMC. The results are shown in Fig.~\ref{CaliforniaBMap} with their values listed in Table \ref{CaliforniaBValues}. Our derived values for B$_{\text{LOS}}$ in the CMC  are not very sensitive to the uncertainties in coordinate and extinction values or to uncertainties in the chemical code input parameters, and their dominant source of uncertainty comes from RM uncertainties.

While there are no Zeeman measurements available for this region to compare with our results, Fig.~\ref{CaliforniaBMap} does exhibit some interesting morphological characteristics. Fig.~\ref{CaliforniaBMap} shows that the magnetic fields on the eastern side of the CMC are pointing away from us, while on the western side they are pointing towards us. This morphology is similar to that seen in OMC-A, and might be an indication of helical magnetic field in this filamentary structure as well.

\subsubsection{Strength and Morphology of B$_{\text{LOS}}$ in the Perseus Molecular Cloud (PMC)}

The PMC is a well-known star forming region at a position of 50$^\circ$<$\alpha (J2000)$<58$^\circ$ and 28$^\circ$<$\delta (J2000)$<34$^\circ$, and at a distance of about 300~pc from the Sun \citep{Ballyetal2008}. To find the proper input physical parameters  to use in our chemical code, we use results found in the literature.  
 \cite{Bachiller1986} study different regions within Perseus and, for the globule L1455 (=B204, B206), they report a  temperature of 12~K. In the position of the NH$_3$ peak they find a density of n(H$_2$) $ \simeq 1.4 \times 10^4 \text{cm}^{-3}$. \cite{BachillerCernicharo1984} mention that B1, has a mean density of n$\succsim 10^3 \text{cm}^{-3}$ and is connected to the rest of the complex with densities of n $\simeq 10^3 \text{cm}^{-3}$. Considering this along with table 2 presented in \cite{Bachiller1986}, we choose 10$^3$ cm$^{-3}$ for the average density and 12~K for the temperature. Additionally, we select a UV field radiation strength of G$_{\circ}$ = 1.0.

\begin{figure*}
\centering
\includegraphics[scale=0.7, trim={1.1cm 4cm 4.1cm 1.8cm},clip]{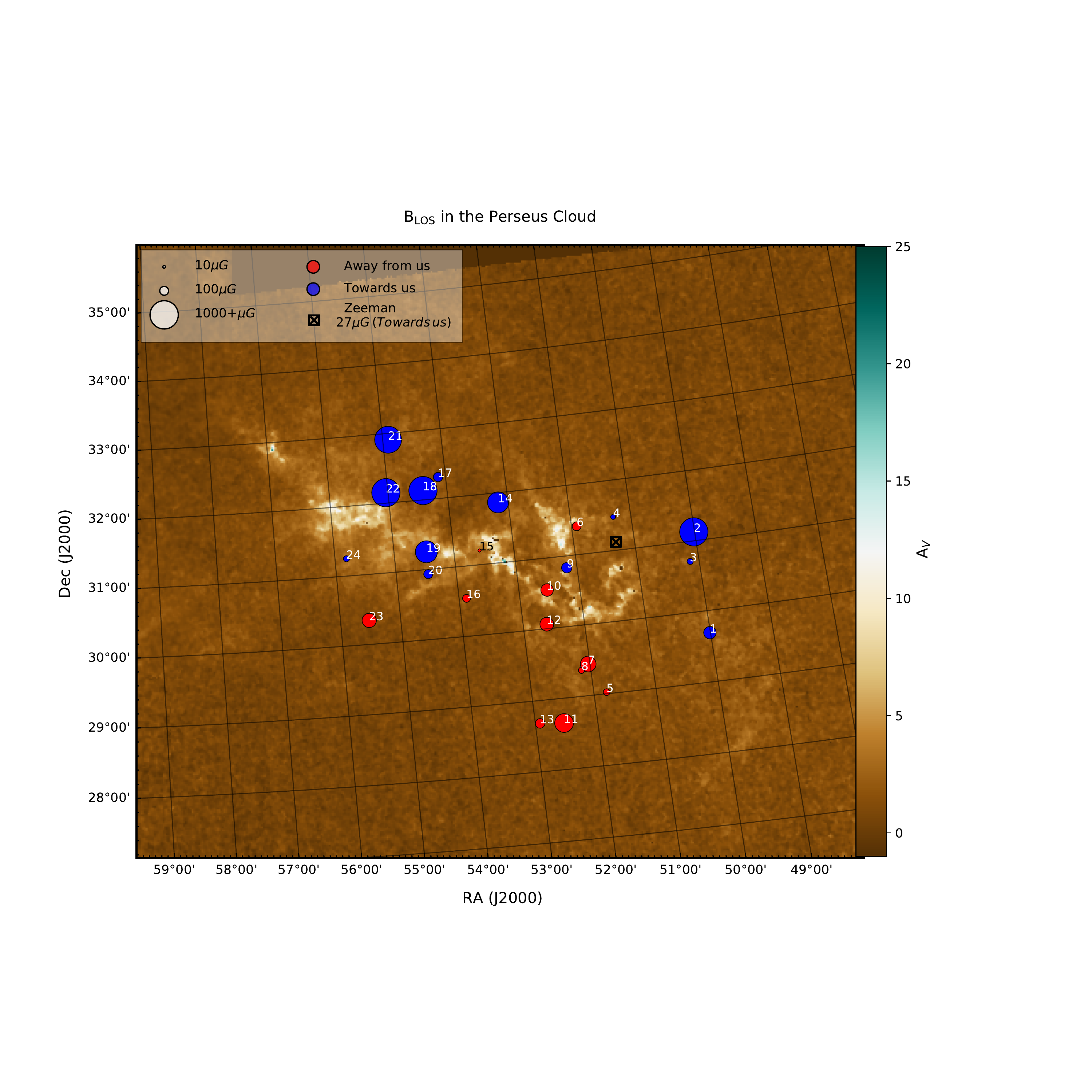}
\caption{\textbf{B}$_{\text{LOS}}$ in the Perseus MC. Blue (red) circles show magnetic fields toward us (away from us). The size of the circles indicate the magnitude of magnetic field. Black square shows the location of the available Zeeman measurements. Color image is the extinction map (A$_{\text{V}}$).}
\label{PerseusBMap}
\end{figure*}

There are several Zeeman measurements available in the well-known B1 molecular core in the PMC  \citep{Goodman1989, Crutcheretal1993, Verschuur1996}, which suggest small magnetic fields. For the B1 region  ($\alpha (J2000)\simeq 51.32^{\circ}, \delta(J2000) \simeq 31.12^{\circ} $), \cite{Goodman1989} obtain a magnetic field of $+27 \pm 4~\mu$G, and \cite{Crutcheretal1993} report $+19.1 \pm 3.9~\mu$G. For the same position \cite{Verschuur1996} finds a magnetic field of $+16.7 \pm 8.9~\mu$G using the 1665 MHz OH line and $-6.2 \pm 8.5~\mu$G using the 1667 OH line. Our closest point to this location is point 4, in Fig.~\ref{PerseusBMap} and Table~\ref{PerseusBValues}, with a value of $+32 \pm 101~\mu$G. Our result is in agreement with all of these reported Zeeman measurements. The  main source of uncertainty of the magnetic field strength using our method is due to uncertainties in the RMs in the TSS09 catalog.

Fig.~\ref{PerseusBMap} seems to suggest that the  magnetic fields on the southern side of the PMC are pointing away from us whereas, on the northern side, they are pointing towards us, however, more data points would be required to draw any firm conclusions since there is a paucity of RMs on the southern side of the cloud. 

\section{Discussion}

\subsection{Decreasing the Uncertainties in \blos }
\label{errordiscussion}
As mentioned previously, our derived magnetic field strengths (see Tables \ref{OrionABValues}, \ref{OrionBBValues}, \ref{CaliforniaBValues}, and \ref{PerseusBValues}) 
often have relatively large uncertainties and, in some cases, the error bars are larger than the tabulated value of \blos .  As mentioned in Sec. \ref{erroranalysis}, the dominant source of errors in our method are the errors of the RMs as tabulated in TSS09.  

The RMs of the TSS09 were calculated using two frequencies in combination with the fractional depolarization as a function of rotation measure.  Errors in the calculated RM could be reduced by re-observing the same sources (in addition to more sources) with new generation radio telescopes such as the Low Frequency Array (LOFAR: a square kilometer Array low pathfinder). For example, in their Table 1, \cite{VanEcketal2017} compare their RM results using LOFAR with the TSS09 catalog.  While the absolute values are in good agreement, the RM uncertainties presented in \cite{VanEcketal2017} (0.05~rad~m$^{-2}$) are significantly smaller than those in TSS09 catalog (10~rad~m$^{-2}$). These reductions in RM uncertainties can accordingly improve the error bars associated with our procedure. For example, for point 4 in OMC-B in Table~\ref{OrionBBValues}, if we hold all other values constant and change the RM uncertainty to 0.05~rad~m$^{-2}$, the final B$_{\text{LOS}}$ would be $122 \pm 50~\mu$G, instead of the currently tabulated $122 \pm 125~\mu$G. In addition, in this ``new and improved'' \blos value of $122 \pm 50~\mu$G, the largest source of error is now from \rmref.  Errors in \rmref\  could be reduced by improved sensitivity RM observations which  could provide additional \rmoff\ data points to be used in the calculation of \rmref .  Since the error in \rmref\ is a standard deviation, with additional points, Poisson statistics should decrease its error.

Even with the current uncertainties in \rmon, we can improve the robustness of our results by removing from consideration any position that has an uncertainty greater than 100\% of the calculated \blos\ value.  Tables \ref{OrionABValuesModif}, \ref{CaliforniaBValuesModif}, and \ref{PerseusBValuesModif} are subsets of Tables \ref{OrionABValues}, \ref{CaliforniaBValues}, and \ref{PerseusBValues} which contain only the points with error bars less than 100\% of the magnetic field strength.  Although the uncertainty in the absolute value of \blos\ of any point may still be relatively high, the \emph{direction} of the magnetic field for the points in these tables is fixed. Therefore, these data can still provide us with insight into the large-scale magnetic field morphology in MCs.  This will be discussed in Sec. \ref{HelicalB} below.

\subsection{Magnetic Field Morphology: Evidence for Helical Fields?}
\label{HelicalB}

In OMC-A, Fig.~\ref{OrionBottomAMap}, suggests that \blos\ on the east side of OMC-A tends to point away from us, whereas on the west side it tends to point towards us. This holds true even if we only use the \blos\ values listed in Table \ref{OrionABValuesModif}, which have error bars small enough that the magnetic field direction is fixed. In fact, our interpretation is more robust using the data in Table \ref{OrionABValuesModif}, since points 21, 28, and 30 on the east side of the cloud and 11, and 16 on the west side of the cloud are removed.  Removing these points strengthens the perceived large-scale pattern of the magnetic field by reducing the number of positions with opposing \blos\ directions.

This magnetic field configuration has been previously observed  in OMC-A \citep{Heiles1997}, and interpreted as a helical magnetic field wrapping around the cloud \citep{Heiles1987}. Other observations,  have also been indirectly interpreted as indications of a helical magnetic field structure \citep[e.g.][]{JohnstonBally1999, Hoqetal2017, Matthews2001, Contrerasetal2013, StutzGould2016}.  For example, by using the Virial mass per length obtained by \cite{FiegePudritzI2000} for a cylindrical filament threaded by a helical magnetic field, \cite{Buckleetal2012} show that the integral shaped filament in OMC-A  is too massive for thermal or turbulent support.  Thus, they suggest that the mass and morphology of the integral shaped filament (a small region within our OMC-A map) is consistent with a Virial model of a filamentary cloud threaded by a helical magnetic field.

In the CMC (Fig.~\ref{CaliforniaBMap}) and PMC (Fig.~\ref{PerseusBMap}), a first glance at the data seem to suggest the presence of a helical magnetic field.  In the CMC this holds true even if we only use the data in  Table \ref{CaliforniaBValuesModif} (with error bars less than 100\% of the \blos\ value).  In the PMC,  if we only use the data in Table~\ref{PerseusBValuesModif}, on the north side of the cloud the remaining points are primarily towards us, but there are too few observations on the southern side of the cloud to truly infer anything about the magnetic field geometry.   

This type of magnetic field geometry is also predicted or investigated by a number of numerical simulations or theoretical analysis \citep{ShibataMatsumoto1991, FiegePudritz1999, FiegePudritz1999Conf,FiegePudritzI2000,FiegePudritzII2000, FiegePudritz2000APJ, SchleicherStutz2017, Nakamuraetal1993, Matsumoto1994, Hanawaetal1993}. \cite{ShibataMatsumoto1991} study the entire Orion Cloud Complex ($\simeq$ 100~pc) and find in their simulations that helically twisted magnetic flux tubes are generated. In addition, \cite{FiegePudritzI2000} and \cite{FiegePudritzII2000} study the fragmentation length-scale, stability, density profile, and mass per length of filamentary MCs and, based on observational constraints, they suggest that many filamentary clouds are likely wrapped by helical magnetic fields.

Additional observations with improved sensitivity and an increased number of RM data points would be required to better map the \blos\ morphology in these MCs and confirm or reject our suggestion of helical magnetic field structure. Such observations should be possible with the new/next generation radio telescope facilities (e.g. LOFAR, SKA).  In addition, simulations of MCs with the sizes and physical characteristics of the OMC-A and CMC are required to theoretically connect the results in this paper to the presence of helical fields.  This will be the subject of a future paper (Tahani et al. in prep, b).

A visual comparison of our results with those of Planck (\cite{PlanckXXXV}) suggests that the data are consistent with a helical or toroidal field wrapping the cloud.  We will investigate the 3D structure of the magnetic field in this region by comparing these two data sets in a more quantitative fashion in a future paper (Tahani et al. in prep, a).

We should note that our technique utilizes OFF positions that are distributed randomly, based on lowest extinction values, around each cloud.  However, it is clear from Figs.~\ref{AllTaurusRM} and \ref{AllOrionRM} that the pattern of a sign change from one side of a cloud to another can sometimes be seen in the raw RM data itself.  Therefore, to investigate whether the observed magnetic field morphology is a result of large-scale Galactic effects or due to the cloud itself, we redo our analysis by choosing OFF positions specific to each side of the cloud.  More precisely, to calculate magnetic fields on one side of the cloud, we choose OFF positions that are on the same side.  For example, for OMC-A, to calculate the magnetic fields on the left side of the cloud where the RMs are predominantly blue (positive), we select OFF positions that are also on the left side of the cloud.  We use the same technique for the right side of the cloud, where the RMs are predominantly red (negative).

We implement this method for OMC-A, CMC, and PMC. We find that considering both sides of the cloud separately and obtaining RM$_{\text{ref}}$ for each side result into a \emph{maximum} change of 5.7~rad~m$^{-2}$, 14.4~rad~m$^{-2}$, and 26.3~rad~m$^{-2}$ from the original RM$_{\text{ref}}$ for OMC-A, CMC, and PMC, respectively. For both OMC-A and CMC, this maximum change is within the original value of $\delta (\text{RM}_{\text{ref}})$. Therefore, for these two clouds the original and updated values of RM$_{\text{ref}}$ are indistinguishable within the uncertainties. Consequently, the overall magnetic field morphology (i.e. direction reversals) in OMC-A and the CMC is preserved in our obtained maps, with very minor and negligible differences.

In the PMC, the changes in RM$_{\text{ref}}$ obtained by using the two sides of the cloud separately are not within the uncertainties. Accordingly the overall magnetic field morphology in PMC is not preserved and the resultant map does not suggest a magnetic field reversal from one side of the cloud to the other. However, since we do not suggest a particular morphology for this region due to a lack of points on the southern side of the cloud, this does not change our original conclusion.

We believe that the choice of the OFF positions, for these clouds, does not affect the overall derived magnetic field morphology. The clouds themselves are located at high Galactic latitudes at longitudes towards the Galactic anti-center, but are only 0.5~kpc away. Thus, the Galactic contribution to the RM along the lines-of-sight will be primarily from the halo, which has an electron density and magnetic field strength each of at least an order of magnitude less than that for the disk, making the RM contribution at least two orders of magnitude less than what would be expected from a similar pathlength entirely through the disk. This does not, of course, exclude the possibility of reversals induced by more local phenomena (e.g. supernova remnants), but we have tried to minimize the possible effects of Galactic-scale structure through our choice of clouds.

We also note that using the bilateral method leads to higher standard deviations in RM$_{\text{ref}}$, i.e. higher values of $\delta (\text{RM}_{\text{ref}})$, and therefore higher uncertainties in the resultant magnetic field strengths.  This is entirely due to the fact that by restricting ourselves to half the area, we have fewer OFF positions with which to calculate RM$_{\text{ref}}$ on each side of the cloud. Consequently, since our original method has smaller error bars and no appreciable difference in the derived overall magnetic field morphology,  we believe that our original choice of reference points with random positions around each cloud is the optimum method to use.

It is very likely that in future studies, with more sensitivity, many more RM points will be available to choose from. A larger dataset would provide smaller statistical errors from a sample of location-specific OFF positions.  Therefore, with a larger number of RM points to choose from, it may be preferable to produce RM$_{\text{ref}}$s on different sides of the clouds  to ensure that one is subtracting out any large scale contributions from the Galaxy.

\subsection{Comparison With Previous Measurements: A Cautionary Note}
\label{ZeemanComparison}
As indicated in \cite{Hulletal2017}, the magnetic field strength and orientation may vary significantly as one moves from higher extinction (small scale) regions to lower extinction (larger scale) regions. For this reason, comparisons between Zeeman measurements and our results must be performed with caution, since they might be probing B$_{\text{LOS}}$ from different regions in the MCs. For example, in our technique we assume that B$_{\text{LOS}}$ is constant in every cloud layer. Thus, in higher extinction regions where we are looking through many cloud layers, we are effectively measuring an average B$_{\text{LOS}}$ along the line of sight (since we use the total electron column density and RM in the MC along the line of sight).  In contrast, Zeeman measurements using one particular molecular line tracer may be selectively probing specific regions/depths in the cloud.  This may also be the reason that different Zeeman measurements in the high extinction regions have a large amount of scatter.  If different measurements probe different layers, they may also be probing different magnetic field strengths in those layers.
In regions with lower extinction, where we are looking through fewer cloud layers, the amount of ``smearing'' over the line-of-sight should be diminished and
we should be more accurately probing the true value of B$_{\text{LOS}}$.  Unfortunately, due to the difficulties inherent in the Zeeman measurement technique, there are few Zeeman measurements in the low column density regions of MCs against which to compare our results. 

\begin{figure*}
\centering
\includegraphics[scale=0.7, trim={.1cm .1cm .1cm .1cm},clip]{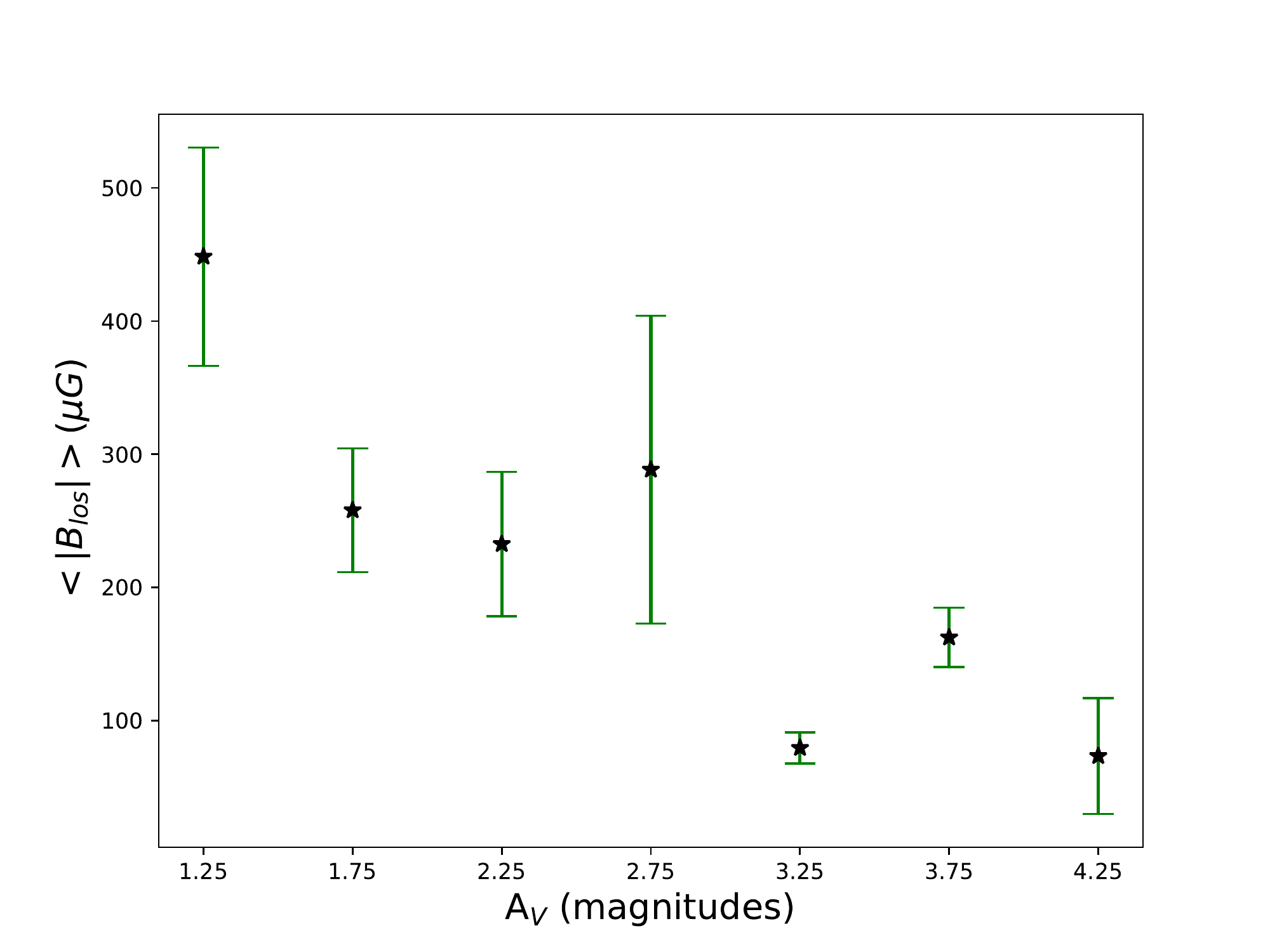}
\caption{The average of absolute value of \blos\ versus extinction, in bins that are 0.5 magnitudes wide in A$_{\text{V}}$. The error bars reflect the standard deviation of \blos\ in each bin. In these data, the average \blos\ appears to decrease with A$_{\text{V}}$.}
\label{BvsAv}
\end{figure*}

Fig.~\ref{BvsAv}, shows the average of absolute value of \blos\ versus extinction, in bins that are 0.5 magnitudes wide in A$_{\text{V}}$. The error bars reflect the standard deviation of $|\text{B}_{\text{LOS}}|$ in each bin. The figure shows a decrease in $<| \text{B}_{\text{LOS}} |>$ with A$_{\text{V}}$, a trend that seems different than that seen in the previous studies \citep[e.g.][]{Lietal2015, Tritsisetal2015}, which explore the magnetic field strength as a function of column density. However, a closer look at Fig.~1 of \cite{Lietal2015} shows that in extinction range of our data points (1 to 4.5), one cannot find a particular trend, within their plotted uncertainties. For extinction magnitude higher than 4.5 there are only a few points available, and up to 30 magnitude only two points. These points are sources 13 and 14 in OMC-A with A$_{\text{V}}$ of 19.56 and 21.47 with \blos\ of $-23 \pm 38$ and $15 \pm 36$, respectively. The points with A$_{\text{V}}$ higher than 30 are sources 1 and 2 in OMC-B, with extinction of 37.36 for both, and \blos\ of $-119 \pm 25$ and $-129 \pm 28$.

The interpretation of our results in Fig.~\ref{BvsAv} should be treated with caution.  Since we are looking through many different cloud layers in the highest A$_{\text{V}}$ regions, and each layer may have a different value of \blos , we are essentially providing an average of  \blos\ through the cloud.  This averaging effect may artificially suppress the  measured value of \blos\  in the highest column density regions, less than that in the low column density regions where there are fewer layers over which to average.

\section{Conclusions}
\label{conclusion}

We present a new method to measure the line-of-sight magnetic field (\blos ) in molecular clouds.  Our technique uses the rotation measures of polarized sources from the catalog of \citet[]{Tayloretal2009} that are located behind, and nearby,  molecular clouds. Using these rotation measures, along with an estimate of electron density determined from extinction maps from \cite{Kainulainenetal2009} and a chemical model, we estimate \blos\ in and around molecular clouds.

We apply our method to four test clouds: the Orion A \&  B cloud complexes, the California molecular cloud, and the Perseus molecular cloud and find good agreement for \blos\ (both in magnitude and direction) with estimates from a limited number of Zeeman measurements in these same regions. For example, in Orion A we calculate \blos $= -23\pm38~\mu$G and $+15\pm36~\mu$G at two positions near the Zeeman measurements.  In the Orion B complex we also  find two rotation measure near the reported Zeeman measurements  with calculated \blos $= -119\pm25~\mu$G and $-129\pm28~\mu$G respectively. In Perseus, our calculated \blos\ at a position nearest the Zeeman measurement is $+32\pm101~\mu$G.
 
The advantage of our method over the traditional Zeeman approach is that we can use the plethora of rotation measures made across the Galaxy to also map the line-of-sight morphology of the magnetic field over large-scales in molecular clouds.  Using this technique, we find that the large-scale morphology of \blos\ in the Orion A complex and the California cloud is suggestive of  helical fields wrapping these clouds.
 Combined with plane-of-the-sky maps of the magnetic field strength and morphology from dust polarization maps, our technique provides a way to determine the true, 3-dimensional structure of the magnetic fields in and around molecular clouds (Tahani et al. 2018a, in prep).

We believe that our method holds great promise for future studies of the large-scale magnetic field morphology in molecular clouds for two reasons.  First,   the magnetic field strengths and directions we calculate are in good qualitative agreement with Zeeman measurements. Second, the inference of helical magnetic field geometries holds true even when we only consider positions with error bars small enough that the \emph{direction} of \blos\ is fixed.

\begin{acknowledgements}
We thank the anonymous referee for his/her insightful comments that helped improve this paper. We would like to thank Tim Robishaw for helpful discussions on Zeeman measurements and their directions. MT acknowledges Eyes High International Doctoral Scholarship. This project has received funding from the European Union's Horizon 2020 research and innovation program under grant agreement No~639459 (PROMISE). We have used LATEX, Python and its associated libraries, PyCharm, SAO Image DS9, and C++ for this work.
\end{acknowledgements}

\bibliographystyle{aa} 
\bibliography{bibt}

\begin{thebibliography}{109}
\expandafter\ifx\csname natexlab\endcsname\relax\def\natexlab#1{#1}\fi

\bibitem[{{Andersson} {et~al.}(2015){Andersson}, {Lazarian}, \&
  {Vaillancourt}}]{Anderssonetal2015}
{Andersson}, B.-G., {Lazarian}, A., \& {Vaillancourt}, J.~E. 2015, \araa, 53,
  501

\bibitem[{{Bachiller} \& {Cernicharo}(1984)}]{BachillerCernicharo1984}
{Bachiller}, R. \& {Cernicharo}, J. 1984, \aap, 140, 414

\bibitem[{{Bachiller} \& {Cernicharo}(1986)}]{Bachiller1986}
{Bachiller}, R. \& {Cernicharo}, J. 1986, \aap, 168, 262

\bibitem[{{Bally} {et~al.}(1991){Bally}, {Langer}, \& {Liu}}]{Ballyetal1991}
{Bally}, J., {Langer}, W.~D., \& {Liu}, W. 1991, \apj, 383, 645

\bibitem[{{Bally} {et~al.}(2008){Bally}, {Walawender}, {Johnstone}, {Kirk}, \&
  {Goodman}}]{Ballyetal2008}
{Bally}, J., {Walawender}, J., {Johnstone}, D., {Kirk}, H., \& {Goodman}, A.
  2008, in Handbook of Star Forming Regions, Volume I, ed. B.~{Reipurth}
  (Astronomical Society of the Pacific), 308

\bibitem[{{Bergin} {et~al.}(1995){Bergin}, {Langer}, \&
  {Goldsmith}}]{BerginLanger1995}
{Bergin}, E.~A., {Langer}, W.~D., \& {Goldsmith}, P.~F. 1995, \apj, 441, 222

\bibitem[{{Bergin} {et~al.}(1999){Bergin}, {Plume}, {Williams}, \&
  {Myers}}]{Berginetal1999}
{Bergin}, E.~A., {Plume}, R., {Williams}, J.~P., \& {Myers}, P.~C. 1999, \apj,
  512, 724

\bibitem[{{Bohlin} {et~al.}(1977){Bohlin}, {Savage}, \&
  {Drake}}]{Bohlinetal1977}
{Bohlin}, R.~C., {Savage}, B.~D., \& {Drake}, J.~F. 1977, NASA STI/Recon
  Technical Report N, 78

\bibitem[{{Brown} {et~al.}(2007){Brown}, {Haverkorn}, {Gaensler}, {Taylor},
  {Bizunok}, {McClure-Griffiths}, {Dickey}, \& {Green}}]{Brownetal2007}
{Brown}, J.~C., {Haverkorn}, M., {Gaensler}, B.~M., {et~al.} 2007, \apj, 663,
  258

\bibitem[{{Brown} {et~al.}(2008){Brown}, {Stil}, \& {Landecker}}]{Brown2008}
{Brown}, J.~C., {Stil}, J.~M., \& {Landecker}, T.~L. 2008, Physics in Canada,
  64

\bibitem[{{Buckle} {et~al.}(2012){Buckle}, {Davis}, {Francesco}, {Graves},
  {Nutter}, {Richer}, {Roberts}, {Ward-Thompson}, {White}, {Brunt}, {Butner},
  {Cavanagh}, {Chrysostomou}, {Curtis}, {Duarte-Cabral}, {Etxaluze}, {Fich},
  {Friberg}, {Friesen}, {Fuller}, {Greaves}, {Hatchell}, {Hogerheijde},
  {Johnstone}, {Matthews}, {Matthews}, {Rawlings}, {Sadavoy}, {Simpson},
  {Tothill}, {Tsamis}, {Viti}, {Wouterloot}, \& {Yates}}]{Buckleetal2012}
{Buckle}, J.~V., {Davis}, C.~J., {Francesco}, J.~D., {et~al.} 2012, \mnras,
  422, 521

\bibitem[{{Castets} {et~al.}(1990){Castets}, {Duvert}, {Dutrey}, {Bally},
  {Langer}, \& {Wilson}}]{Castetsetal1990}
{Castets}, A., {Duvert}, G., {Dutrey}, A., {et~al.} 1990, \aap, 234, 469

\bibitem[{{Clark} \& {Glover}(2014)}]{ClarkGlover2014}
{Clark}, P.~C. \& {Glover}, S.~C.~O. 2014, \mnras, 444, 2396

\bibitem[{{Condon} {et~al.}(1998){Condon}, {Cotton}, {Greisen}, {Yin},
  {Perley}, {Taylor}, \& {Broderick}}]{Condon1998}
{Condon}, J.~J., {Cotton}, W.~D., {Greisen}, E.~W., {et~al.} 1998, \aj, 115,
  1693

\bibitem[{{Contreras} {et~al.}(2013){Contreras}, {Rathborne}, \&
  {Garay}}]{Contrerasetal2013}
{Contreras}, Y., {Rathborne}, J., \& {Garay}, G. 2013, \mnras, 433, 251

\bibitem[{{Cox}(2005)}]{Cox2005}
{Cox}, D.~P. 2005, \araa, 43, 337

\bibitem[{{Crutcher}(2005)}]{Crutcher2005}
{Crutcher}, R. 2005, in The Magnetized Plasma in Galaxy Evolution, ed. K.~T.
  {Chyzy}, K.~{Otmianowska-Mazur}, M.~{Soida}, \& R.-J. {Dettmar}, 103--110

\bibitem[{{Crutcher}(1999)}]{Crutcher1999APJ520}
{Crutcher}, R.~M. 1999, \apj, 520, 706

\bibitem[{{Crutcher} {et~al.}(1999{\natexlab{a}}){Crutcher}, {Roberts},
  {Troland}, \& {Goss}}]{Crutcheretal1999APJ515}
{Crutcher}, R.~M., {Roberts}, D.~A., {Troland}, T.~H., \& {Goss}, W.~M.
  1999{\natexlab{a}}, \apj, 515, 275

\bibitem[{{Crutcher} \& {Troland}(2008)}]{CrutcherTroland2008}
{Crutcher}, R.~M. \& {Troland}, T.~H. 2008, \apj, 685, 281

\bibitem[{{Crutcher} {et~al.}(1993){Crutcher}, {Troland}, {Goodman}, {Heiles},
  {Kazes}, \& {Myers}}]{Crutcheretal1993}
{Crutcher}, R.~M., {Troland}, T.~H., {Goodman}, A.~A., {et~al.} 1993, \apj,
  407, 175

\bibitem[{{Crutcher} {et~al.}(1996){Crutcher}, {Troland}, {Lazareff}, \&
  {Kazes}}]{Crutcheretal1996}
{Crutcher}, R.~M., {Troland}, T.~H., {Lazareff}, B., \& {Kazes}, I. 1996, \apj,
  456, 217

\bibitem[{{Crutcher} {et~al.}(1999{\natexlab{b}}){Crutcher}, {Troland},
  {Lazareff}, {Paubert}, \& {Kaz{\`e}s}}]{Crutcher1999APJ514}
{Crutcher}, R.~M., {Troland}, T.~H., {Lazareff}, B., {Paubert}, G., \&
  {Kaz{\`e}s}, I. 1999{\natexlab{b}}, \apjl, 514, L121

\bibitem[{{Crutcher} {et~al.}(2010){Crutcher}, {Wandelt}, {Heiles},
  {Falgarone}, \& {Troland}}]{Crutcher2010}
{Crutcher}, R.~M., {Wandelt}, B., {Heiles}, C., {Falgarone}, E., \& {Troland},
  T.~H. 2010, \apj, 725, 466

\bibitem[{{Dutrey} {et~al.}(1993){Dutrey}, {Duvert}, {Castets}, {Langer},
  {Bally}, \& {Wilson}}]{Dutreyetal1993}
{Dutrey}, A., {Duvert}, G., {Castets}, A., {et~al.} 1993, \aap, 270, 468

\bibitem[{{Everett} \& {Zweibel}(2011)}]{EverettZwibel2011}
{Everett}, J.~E. \& {Zweibel}, E.~G. 2011, \apj, 739, 60

\bibitem[{{Falgarone} {et~al.}(2008){Falgarone}, {Troland}, {Crutcher}, \&
  {Paubert}}]{Falgaroneetal2008}
{Falgarone}, E., {Troland}, T.~H., {Crutcher}, R.~M., \& {Paubert}, G. 2008,
  \aap, 487, 247

\bibitem[{{Fiege} \& {Pudritz}(1999{\natexlab{a}})}]{FiegePudritz1999}
{Fiege}, J.~D. \& {Pudritz}, R.~E. 1999{\natexlab{a}}, ArXiv Astrophysics
  e-prints [\eprint{astro-ph/9905148}]

\bibitem[{{Fiege} \& {Pudritz}(1999{\natexlab{b}})}]{FiegePudritz1999Conf}
{Fiege}, J.~D. \& {Pudritz}, R.~E. 1999{\natexlab{b}}, in Astronomical Society
  of the Pacific Conference Series, Vol. 168, New Perspectives on the
  Interstellar Medium, ed. A.~R. {Taylor}, T.~L. {Landecker}, \& G.~{Joncas},
  248

\bibitem[{{Fiege} \& {Pudritz}(2000{\natexlab{a}})}]{FiegePudritzI2000}
{Fiege}, J.~D. \& {Pudritz}, R.~E. 2000{\natexlab{a}}, \mnras, 311, 85

\bibitem[{{Fiege} \& {Pudritz}(2000{\natexlab{b}})}]{FiegePudritzII2000}
{Fiege}, J.~D. \& {Pudritz}, R.~E. 2000{\natexlab{b}}, \mnras, 311, 105

\bibitem[{{Fiege} \& {Pudritz}(2000{\natexlab{c}})}]{FiegePudritz2000APJ}
{Fiege}, J.~D. \& {Pudritz}, R.~E. 2000{\natexlab{c}}, \apj, 544, 830

\bibitem[{{Flower} {et~al.}(2007){Flower}, {Pineau Des For{\^e}ts}, \&
  {Walmsley}}]{Floweretal2007}
{Flower}, D.~R., {Pineau Des For{\^e}ts}, G., \& {Walmsley}, C.~M. 2007, \aap,
  474, 923

\bibitem[{{Gibson} {et~al.}(2009){Gibson}, {Plume}, {Bergin}, {Ragan}, \&
  {Evans}}]{Gibsonetal2009}
{Gibson}, D., {Plume}, R., {Bergin}, E., {Ragan}, S., \& {Evans}, N. 2009,
  \apj, 705, 123

\bibitem[{{Glover} \& {Clark}(2012)}]{GloverClark2012}
{Glover}, S.~C.~O. \& {Clark}, P.~C. 2012, \mnras, 426, 377

\bibitem[{{Glover} {et~al.}(2010){Glover}, {Federrath}, {Mac Low}, \&
  {Klessen}}]{Gloveretal2010}
{Glover}, S.~C.~O., {Federrath}, C., {Mac Low}, M.-M., \& {Klessen}, R.~S.
  2010, \mnras, 404, 2

\bibitem[{{Goldsmith} {et~al.}(2008){Goldsmith}, {Heyer}, {Narayanan}, {Snell},
  {Li}, \& {Brunt}}]{Goldsmith2008}
{Goldsmith}, P.~F., {Heyer}, M., {Narayanan}, G., {et~al.} 2008, \apj, 680, 428

\bibitem[{{Goodman} {et~al.}(1989){Goodman}, {Crutcher}, {Heiles}, {Myers}, \&
  {Troland}}]{Goodman1989}
{Goodman}, A.~A., {Crutcher}, R.~M., {Heiles}, C., {Myers}, P.~C., \&
  {Troland}, T.~H. 1989, \apjl, 338, L61

\bibitem[{{Han} {et~al.}(2006){Han}, {Manchester}, {Lyne}, {Qiao}, \& {van
  Straten}}]{Hanetal2006}
{Han}, J.~L., {Manchester}, R.~N., {Lyne}, A.~G., {Qiao}, G.~J., \& {van
  Straten}, W. 2006, \apj, 642, 868

\bibitem[{{Hanawa} {et~al.}(1993){Hanawa}, {Nakamura}, {Matsumoto}, {Nakano},
  {Tatematsu}, {Umemoto}, {Kameya}, {Hirano}, {Hasegawa}, {Kaifu}, \&
  {Yamamoto}}]{Hanawaetal1993}
{Hanawa}, T., {Nakamura}, F., {Matsumoto}, T., {et~al.} 1993, \apjl, 404, L83

\bibitem[{{Harrison} {et~al.}(2013){Harrison}, {Faure}, \&
  {Tennyson}}]{Harrisonetal2013}
{Harrison}, S., {Faure}, A., \& {Tennyson}, J. 2013, \mnras, 435, 3541

\bibitem[{{Harvey} {et~al.}(2013){Harvey}, {Fallscheer}, {Ginsburg}, {Terebey},
  {Andr{\'e}}, {Bourke}, {Di Francesco}, {K{\"o}nyves}, {Matthews}, \&
  {Peterson}}]{Harveyetal2013}
{Harvey}, P.~M., {Fallscheer}, C., {Ginsburg}, A., {et~al.} 2013, \apj, 764,
  133

\bibitem[{{Hasegawa} \& {Herbst}(1993)}]{1993HasegawaHerbst}
{Hasegawa}, T.~I. \& {Herbst}, E. 1993, \mnras, 263, 589

\bibitem[{{Heiles}(1987)}]{Heiles1987}
{Heiles}, C. 1987, in Astrophysics and Space Science Library, Vol. 134,
  Interstellar Processes, ed. D.~J. {Hollenbach} \& H.~A. {Thronson}, Jr.,
  171--194

\bibitem[{{Heiles}(1997)}]{Heiles1997}
{Heiles}, C. 1997, \apjs, 111, 245

\bibitem[{{Hollenbach} \& {Tielens}(1995)}]{HollenbachTielens1995}
{Hollenbach}, D.~J. \& {Tielens}, A.~G.~G.~M. 1995, in Lecture Notes in
  Physics, Berlin Springer Verlag, Vol. 459, The Physics and Chemistry of
  Interstellar Molecular Clouds, ed. G.~{Winnewisser} \& G.~C. {Pelz}, 164--174

\bibitem[{{Hollenbach} \& {Tielens}(1997)}]{HollenbachTielens1997}
{Hollenbach}, D.~J. \& {Tielens}, A.~G.~G.~M. 1997, \araa, 35, 179

\bibitem[{{Hollenbach} \& {Tielens}(1999)}]{HollenbachTielens1999}
{Hollenbach}, D.~J. \& {Tielens}, A.~G.~G.~M. 1999, Reviews of Modern Physics,
  71, 173

\bibitem[{{Hoq} {et~al.}(2017){Hoq}, {Clemens}, {Guzm{\'a}n}, \&
  {Cashman}}]{Hoqetal2017}
{Hoq}, S., {Clemens}, D.~P., {Guzm{\'a}n}, A.~E., \& {Cashman}, L.~R. 2017,
  \apj, 836, 199

\bibitem[{{Hull} {et~al.}(2017){Hull}, {Girart}, {Tychoniec}, {Rao},
  {Cort{\'e}s}, {Pokhrel}, {Zhang}, {Houde}, {Dunham}, {Kristensen}, {Lai},
  {Li}, \& {Plambeck}}]{Hulletal2017}
{Hull}, C.~L.~H., {Girart}, J.~M., {Tychoniec}, {\L}., {et~al.} 2017, ArXiv
  e-prints [\eprint[arXiv]{1707.03827}]

\bibitem[{{Johnstone} \& {Bally}(1999{\natexlab{a}})}]{JohnstonBally1999}
{Johnstone}, D. \& {Bally}, J. 1999{\natexlab{a}}, \apjl, 510, L49

\bibitem[{{Johnstone} \& {Bally}(1999{\natexlab{b}})}]{JohnstoneBally1999a}
{Johnstone}, D. \& {Bally}, J. 1999{\natexlab{b}}, \apjl, 510, L49

\bibitem[{{Johnstone} \& {Bally}(1999{\natexlab{c}})}]{JohnstoneBally1999b}
{Johnstone}, D. \& {Bally}, J. 1999{\natexlab{c}}, in The Physics and Chemistry
  of the Interstellar Medium, ed. V.~{Ossenkopf}, J.~{Stutzki}, \&
  G.~{Winnewisser}

\bibitem[{{Johnstone} \& {Bally}(2006)}]{JohnstoneBally2006}
{Johnstone}, D. \& {Bally}, J. 2006, \apj, 653, 383

\bibitem[{{Kainulainen} {et~al.}(2009){Kainulainen}, {Beuther}, {Henning}, \&
  {Plume}}]{Kainulainenetal2009}
{Kainulainen}, J., {Beuther}, H., {Henning}, T., \& {Plume}, R. 2009, \aap,
  508, L35

\bibitem[{{Khesali} {et~al.}(2014){Khesali}, {Kokabi}, {Faghei}, \&
  {Nejad-Asghar}}]{Khesalietal2014}
{Khesali}, A., {Kokabi}, K., {Faghei}, K., \& {Nejad-Asghar}, M. 2014, Research
  in Astronomy and Astrophysics, 14, 66

\bibitem[{{Killeen} {et~al.}(1992){Killeen}, {Lo}, \&
  {Crutcher}}]{Killeeetal1992}
{Killeen}, N.~E.~B., {Lo}, K.~Y., \& {Crutcher}, R. 1992, \apj, 385, 585

\bibitem[{{Kirk} {et~al.}(2015){Kirk}, {Klassen}, {Pudritz}, \&
  {Pillsworth}}]{Kirketal2015}
{Kirk}, H., {Klassen}, M., {Pudritz}, R., \& {Pillsworth}, S. 2015, \apj, 802,
  75

\bibitem[{{Klassen} {et~al.}(2017){Klassen}, {Pudritz}, \&
  {Kirk}}]{Klessenetal2017}
{Klassen}, M., {Pudritz}, R.~E., \& {Kirk}, H. 2017, \mnras, 465, 2254

\bibitem[{{Kong} {et~al.}(2015){Kong}, {Lada}, {Lada},
  {Rom{\'a}n-Z{\'u}{\~n}iga}, {Bieging}, {Lombardi}, {Forbrich}, \&
  {Alves}}]{Kongetal2015}
{Kong}, S., {Lada}, C.~J., {Lada}, E.~A., {et~al.} 2015, \apj, 805, 58

\bibitem[{{Kounkel} {et~al.}(2017){Kounkel}, {Hartmann}, {Loinard},
  {Ortiz-Le{\'o}n}, {Mioduszewski}, {Rodr{\'{\i}}guez}, {Dzib}, {Torres},
  {Pech}, {Galli}, {Rivera}, {Boden}, {Evans}, {Brice{\~n}o}, \&
  {Tobin}}]{Kounkeletal2017}
{Kounkel}, M., {Hartmann}, L., {Loinard}, L., {et~al.} 2017, \apj, 834, 142

\bibitem[{{Lada} {et~al.}(2009){Lada}, {Lombardi}, \& {Alves}}]{Ladaetal2009}
{Lada}, C.~J., {Lombardi}, M., \& {Alves}, J.~F. 2009, \apj, 703, 52

\bibitem[{{Le Teuff} {et~al.}(2000){Le Teuff}, {Millar}, \&
  {Markwick}}]{LeTeuffetal1999}
{Le Teuff}, Y.~H., {Millar}, T.~J., \& {Markwick}, A.~J. 2000, \aaps, 146, 157

\bibitem[{{Li} {et~al.}(2014){Li}, {Goodman}, {Sridharan}, {Houde}, {Li},
  {Novak}, \& {Tang}}]{Lietal2014}
{Li}, H.-B., {Goodman}, A., {Sridharan}, T.~K., {et~al.} 2014, Protostars and
  Planets VI, 101

\bibitem[{{Li} {et~al.}(2015){Li}, {McKee}, \& {Klein}}]{Lietal2015}
{Li}, P.~S., {McKee}, C.~F., \& {Klein}, R.~I. 2015, \mnras, 452, 2500

\bibitem[{{Lombardi}(2009)}]{Lombardi2009Nicest}
{Lombardi}, M. 2009, \aap, 493, 735

\bibitem[{{Lombardi} {et~al.}(2010){Lombardi}, {Lada}, \&
  {Alves}}]{Lombardietal2010}
{Lombardi}, M., {Lada}, C.~J., \& {Alves}, J. 2010, \aap, 512, A67

\bibitem[{{Matsumoto} {et~al.}(1994){Matsumoto}, {Nakamura}, \&
  {Hanawa}}]{Matsumoto1994}
{Matsumoto}, T., {Nakamura}, F., \& {Hanawa}, T. 1994, \pasj, 46, 243

\bibitem[{{Matthews} {et~al.}(2001){Matthews}, {Wilson}, \&
  {Fiege}}]{Matthews2001}
{Matthews}, B.~C., {Wilson}, C.~D., \& {Fiege}, J.~D. 2001, \apj, 562, 400

\bibitem[{{McKee} \& {Ostriker}(2007)}]{MckeeOstriker2007}
{McKee}, C.~F. \& {Ostriker}, E.~C. 2007, \araa, 45, 565

\bibitem[{Media~Relations(2006)}]{PressRelease}
Media~Relations, R.~S. 2006, astronomers find magnetic slinky in orion,
  \url{www.berkeley.edu/news/media/releases/2006/01/12_helical.shtml}

\bibitem[{{Mitchell} {et~al.}(2001){Mitchell}, {Johnstone},
  {Moriarty-Schieven}, {Fich}, \& {Tothill}}]{Mitchelletal2001}
{Mitchell}, G.~F., {Johnstone}, D., {Moriarty-Schieven}, G., {Fich}, M., \&
  {Tothill}, N.~F.~H. 2001, \apj, 556, 215

\bibitem[{{Morlino} \& {Gabici}(2015)}]{MorlinoGabici2015}
{Morlino}, G. \& {Gabici}, S. 2015, \mnras, 451, L100

\bibitem[{{Morlino} {et~al.}(2015){Morlino}, {Gabici}, \&
  {Krause}}]{Morlinoetal2015}
{Morlino}, G., {Gabici}, S., \& {Krause}, J. 2015, ArXiv e-prints
  [\eprint[arXiv]{1509.05128}]

\bibitem[{{Nakamura} {et~al.}(1993){Nakamura}, {Hanawa}, \&
  {Nakano}}]{Nakamuraetal1993}
{Nakamura}, F., {Hanawa}, T., \& {Nakano}, T. 1993, \pasj, 45, 551

\bibitem[{{Ordog} {et~al.}(2017){Ordog}, {Brown}, {Kothes}, \&
  {Landecker}}]{Ordogetal2017}
{Ordog}, A., {Brown}, J.~C., {Kothes}, R., \& {Landecker}, T.~L. 2017, \aap,
  603, A15

\bibitem[{{Padovani} \& {Galli}(2013)}]{PadovaniGalli2013}
{Padovani}, M. \& {Galli}, D. 2013, in Astrophysics and Space Science
  Proceedings, Vol.~34, Cosmic Rays in Star-Forming Environments, ed. D.~F.
  {Torres} \& O.~{Reimer}, 61

\bibitem[{{Padovani} {et~al.}(2009){Padovani}, {Galli}, \&
  {Glassgold}}]{Padovanietal2009}
{Padovani}, M., {Galli}, D., \& {Glassgold}, A.~E. 2009, \aap, 501, 619

\bibitem[{{Padovani} {et~al.}(2016){Padovani}, {Marcowith}, {Hennebelle}, \&
  {Ferri{\`e}re}}]{Padovanietal2016}
{Padovani}, M., {Marcowith}, A., {Hennebelle}, P., \& {Ferri{\`e}re}, K. 2016,
  \aap, 590, A8

\bibitem[{{Palmeirim} {et~al.}(2013){Palmeirim}, {Andr{\'e}}, {Kirk},
  {Ward-Thompson}, {Arzoumanian}, {K{\"o}nyves}, {Didelon}, {Schneider},
  {Benedettini}, {Bontemps}, {Di Francesco}, {Elia}, {Griffin}, {Hennemann},
  {Hill}, {Martin}, {Men'shchikov}, {Molinari}, {Motte}, {Nguyen Luong},
  {Nutter}, {Peretto}, {Pezzuto}, {Roy}, {Rygl}, {Spinoglio}, \&
  {White}}]{Palmeirim2013}
{Palmeirim}, P., {Andr{\'e}}, P., {Kirk}, J., {et~al.} 2013, \aap, 550, A38

\bibitem[{{Planck Collaboration} {et~al.}(2016){Planck Collaboration}, {Ade},
  {Aghanim}, {Alves}, {Arnaud}, {Arzoumanian}, {Ashdown}, {Aumont},
  {Baccigalupi}, {Banday}, {Barreiro}, {Bartolo}, {Battaner}, {Benabed},
  {Beno{\^i}t}, {Benoit-L{\'e}vy}, {Bernard}, {Bersanelli}, {Bielewicz},
  {Bock}, {Bonavera}, {Bond}, {Borrill}, {Bouchet}, {Boulanger}, {Bracco},
  {Burigana}, {Calabrese}, {Cardoso}, {Catalano}, {Chiang}, {Christensen},
  {Colombo}, {Combet}, {Couchot}, {Crill}, {Curto}, {Cuttaia}, {Danese},
  {Davies}, {Davis}, {de Bernardis}, {de Rosa}, {de Zotti}, {Delabrouille},
  {Dickinson}, {Diego}, {Dole}, {Donzelli}, {Dor{\'e}}, {Douspis}, {Ducout},
  {Dupac}, {Efstathiou}, {Elsner}, {En{\ss}lin}, {Eriksen}, {Falceta-Gon{\c
  c}alves}, {Falgarone}, {Ferri{\`e}re}, {Finelli}, {Forni}, {Frailis},
  {Fraisse}, {Franceschi}, {Frejsel}, {Galeotta}, {Galli}, {Ganga}, {Ghosh},
  {Giard}, {Gjerl{\o}w}, {Gonz{\'a}lez-Nuevo}, {G{\'o}rski}, {Gregorio},
  {Gruppuso}, {Gudmundsson}, {Guillet}, {Harrison}, {Helou}, {Hennebelle},
  {Henrot-Versill{\'e}}, {Hern{\'a}ndez-Monteagudo}, {Herranz}, {Hildebrandt},
  {Hivon}, {Holmes}, {Hornstrup}, {Huffenberger}, {Hurier}, {Jaffe}, {Jaffe},
  {Jones}, {Juvela}, {Keih{\"a}nen}, {Keskitalo}, {Kisner}, {Knoche}, {Kunz},
  {Kurki-Suonio}, {Lagache}, {Lamarre}, {Lasenby}, {Lattanzi}, {Lawrence},
  {Leonardi}, {Levrier}, {Liguori}, {Lilje}, {Linden-V{\o}rnle},
  {L{\'o}pez-Caniego}, {Lubin}, {Mac{\'{\i}}as-P{\'e}rez}, {Maino},
  {Mandolesi}, {Mangilli}, {Maris}, {Martin}, {Mart{\'{\i}}nez-Gonz{\'a}lez},
  {Masi}, {Matarrese}, {Melchiorri}, {Mendes}, {Mennella}, {Migliaccio},
  {Miville-Desch{\^e}nes}, {Moneti}, {Montier}, {Morgante}, {Mortlock},
  {Munshi}, {Murphy}, {Naselsky}, {Nati}, {Netterfield}, {Noviello}, {Novikov},
  {Novikov}, {Oppermann}, {Oxborrow}, {Pagano}, {Pajot}, {Paladini},
  {Paoletti}, {Pasian}, {Perotto}, {Pettorino}, {Piacentini}, {Piat},
  {Pierpaoli}, {Pietrobon}, {Plaszczynski}, {Pointecouteau}, {Polenta},
  {Ponthieu}, {Pratt}, {Prunet}, {Puget}, {Rachen}, {Reinecke}, {Remazeilles},
  {Renault}, {Renzi}, {Ristorcelli}, {Rocha}, {Rossetti}, {Roudier},
  {Rubi{\~n}o-Mart{\'{\i}}n}, {Rusholme}, {Sandri}, {Santos}, {Savelainen},
  {Savini}, {Scott}, {Soler}, {Stolyarov}, {Sudiwala}, {Sutton}, {Suur-Uski},
  {Sygnet}, {Tauber}, {Terenzi}, {Toffolatti}, {Tomasi}, {Tristram}, {Tucci},
  {Umana}, {Valenziano}, {Valiviita}, {Van Tent}, {Vielva}, {Villa}, {Wade},
  {Wandelt}, {Wehus}, {Ysard}, {Yvon}, \& {Zonca}}]{PlanckXXXV}
{Planck Collaboration}, {Ade}, P.~A.~R., {Aghanim}, N., {et~al.} 2016, \aap,
  586, A138

\bibitem[{{Pudritz} {et~al.}(2014){Pudritz}, {Klassen}, {Kirk}, {Seifried}, \&
  {Banerjee}}]{Pudritzetal2014}
{Pudritz}, R.~E., {Klassen}, M., {Kirk}, H., {Seifried}, D., \& {Banerjee}, R.
  2014, in IAU Symposium, Vol. 302, Magnetic Fields throughout Stellar
  Evolution, ed. P.~{Petit}, M.~{Jardine}, \& H.~C. {Spruit}, 10--20

\bibitem[{{Reich} {et~al.}(2002){Reich}, {F{\"u}rst}, {Reich}, {Wielebinski},
  \& {Wolleben}}]{Reichetal2002}
{Reich}, W., {F{\"u}rst}, E., {Reich}, P., {Wielebinski}, R., \& {Wolleben}, M.
  2002, in American Institute of Physics Conference Series, Vol. 609,
  Astrophysical Polarized Backgrounds, ed. S.~{Cecchini}, S.~{Cortiglioni},
  R.~{Sault}, \& C.~{Sbarra}, 3--8

\bibitem[{{Robishaw}(2008)}]{RobishawThesis}
{Robishaw}, T. 2008, PhD thesis, University of California, Berkeley

\bibitem[{{R{\"o}llig} {et~al.}(2007){R{\"o}llig}, {Abel}, {Bell}, {Bensch},
  {Black}, {Ferland}, {Jonkheid}, {Kamp}, {Kaufman}, {Le Bourlot}, {Le Petit},
  {Meijerink}, {Morata}, {Ossenkopf}, {Roueff}, {Shaw}, {Spaans}, {Sternberg},
  {Stutzki}, {Thi}, {van Dishoeck}, {van Hoof}, {Viti}, \&
  {Wolfire}}]{Rollingetal2007}
{R{\"o}llig}, M., {Abel}, N.~P., {Bell}, T., {et~al.} 2007, \aap, 467, 187

\bibitem[{{Schleicher} \& {Stutz}(2017)}]{SchleicherStutz2017}
{Schleicher}, D.~R.~G. \& {Stutz}, A.~M. 2017, ArXiv e-prints
  [\eprint[arXiv]{1705.06302}]

\bibitem[{{Schnee} {et~al.}(2014){Schnee}, {Mason}, {Di Francesco}, {Friesen},
  {Li}, {Sadavoy}, \& {Stanke}}]{Schneeetal2014}
{Schnee}, S., {Mason}, B., {Di Francesco}, J., {et~al.} 2014, \mnras, 444, 2303

\bibitem[{{Seifried} {et~al.}(2017){Seifried}, {Walch}, {Girichidis}, {Naab},
  {W{\"u}nsch}, {Klessen}, {Glover}, {Peters}, \& {Clark}}]{Seifried2017}
{Seifried}, D., {Walch}, S., {Girichidis}, P., {et~al.} 2017, ArXiv e-prints
  [\eprint[arXiv]{1704.06487}]

\bibitem[{{Shibata} \& {Matsumoto}(1991)}]{ShibataMatsumoto1991}
{Shibata}, K. \& {Matsumoto}, R. 1991, \nat, 353, 633

\bibitem[{{Simard-Normandin} \& {Kronberg}(1980)}]{SimardKronberg1980}
{Simard-Normandin}, M. \& {Kronberg}, P.~P. 1980, \apj, 242, 74

\bibitem[{{Smith} {et~al.}(2014){Smith}, {Glover}, {Clark}, {Klessen}, \&
  {Springel}}]{Smithetal2014}
{Smith}, R.~J., {Glover}, S.~C.~O., {Clark}, P.~C., {Klessen}, R.~S., \&
  {Springel}, V. 2014, \mnras, 441, 1628

\bibitem[{{Stil} \& {Hryhoriw}(2016)}]{StilHryhoriw2016}
{Stil}, J.~M. \& {Hryhoriw}, A. 2016, \apj, 826, 202

\bibitem[{{Stutz} \& {Gould}(2016)}]{StutzGould2016}
{Stutz}, A.~M. \& {Gould}, A. 2016, \aap, 590, A2

\bibitem[{{Sun} {et~al.}(2008){Sun}, {Reich}, {Waelkens}, \&
  {En{\ss}lin}}]{Sunetal2008}
{Sun}, X.~H., {Reich}, W., {Waelkens}, A., \& {En{\ss}lin}, T.~A. 2008, \aap,
  477, 573

\bibitem[{{Taylor} {et~al.}(2009){Taylor}, {Stil}, \&
  {Sunstrum}}]{Tayloretal2009}
{Taylor}, A.~R., {Stil}, J.~M., \& {Sunstrum}, C. 2009, \apj, 702, 1230

\bibitem[{{Taylor}(1997)}]{UncertaintyBook}
{Taylor}, L.~R. 1997, An Introduction to Error Analysis, 2nd edn. (University
  Science Books)

\bibitem[{{Tritsis} {et~al.}(2015){Tritsis}, {Panopoulou}, {Mouschovias},
  {Tassis}, \& {Pavlidou}}]{Tritsisetal2015}
{Tritsis}, A., {Panopoulou}, G.~V., {Mouschovias}, T.~C., {Tassis}, K., \&
  {Pavlidou}, V. 2015, \mnras, 451, 4384

\bibitem[{{Troland} {et~al.}(1986){Troland}, {Crutcher}, \&
  {Kazes}}]{Trolandetal1986}
{Troland}, T.~H., {Crutcher}, R.~M., \& {Kazes}, I. 1986, \apjl, 304, L57

\bibitem[{{Troland} \& {Heiles}(1982)}]{TrolandHeiles1982}
{Troland}, T.~H. \& {Heiles}, C. 1982, \apj, 252, 179

\bibitem[{{Troland} {et~al.}(1989){Troland}, {Heiles}, \&
  {Goss}}]{Trolandetal1989}
{Troland}, T.~H., {Heiles}, C., \& {Goss}, W.~M. 1989, \apj, 337, 342

\bibitem[{{Van Eck} {et~al.}(2011){Van Eck}, {Brown}, {Stil}, {Rae}, {Mao},
  {Gaensler}, {Shukurov}, {Taylor}, {Haverkorn}, {Kronberg}, \&
  {McClure-Griffiths}}]{VanEcketal2011}
{Van Eck}, C.~L., {Brown}, J.~C., {Stil}, J.~M., {et~al.} 2011, \apj, 728, 97

\bibitem[{{Van Eck} {et~al.}(2017){Van Eck}, {Haverkorn}, {Alves}, {Beck}, {de
  Bruyn}, {En{\ss}lin}, {Farnes}, {Ferri{\`e}re}, {Heald}, {Horellou},
  {Horneffer}, {Iacobelli}, {Jeli{\'c}}, {Mart{\'{\i}}-Vidal}, {Mulcahy},
  {Reich}, {R{\"o}ttgering}, {Scaife}, {Schnitzeler}, {Sobey}, \&
  {Sridhar}}]{VanEcketal2017}
{Van Eck}, C.~L., {Haverkorn}, M., {Alves}, M.~I.~R., {et~al.} 2017, \aap, 597,
  A98

\bibitem[{{Van Loo} {et~al.}(2014){Van Loo}, {Keto}, \& {Zhang}}]{VanLoo2014}
{Van Loo}, S., {Keto}, E., \& {Zhang}, Q. 2014, \apj, 789, 37

\bibitem[{{Verschuur}(1996)}]{Verschuur1996}
{Verschuur}, G.~L. 1996, \aj, 112, 2718

\bibitem[{{Willacy} \& {Williams}(1993)}]{WillacyWilliams1993}
{Willacy}, K. \& {Williams}, D.~A. 1993, \mnras, 260, 635

\bibitem[{{Williams} {et~al.}(1998){Williams}, {Bergin}, {Caselli}, {Myers}, \&
  {Plume}}]{Williamsetal1998}
{Williams}, J.~P., {Bergin}, E.~A., {Caselli}, P., {Myers}, P.~C., \& {Plume},
  R. 1998, \apj, 503, 689

\bibitem[{{Wolleben} \& {Reich}(2004{\natexlab{a}})}]{WollebenReichb2004}
{Wolleben}, M. \& {Reich}, W. 2004{\natexlab{a}}, \aap, 427, 537

\bibitem[{{Wolleben} \& {Reich}(2004{\natexlab{b}})}]{WollebenReich2004}
{Wolleben}, M. \& {Reich}, W. 2004{\natexlab{b}}, in The Magnetized
  Interstellar Medium, ed. B.~{Uyaniker}, W.~{Reich}, \& R.~{Wielebinski},
  99--104

\bibitem[{{Yusef-Zadeh} {et~al.}(1997){Yusef-Zadeh}, {Wardle}, \&
  {Parastaran}}]{YusefZadehetal1997}
{Yusef-Zadeh}, F., {Wardle}, M., \& {Parastaran}, P. 1997, \apjl, 475, L119

\end{thebibliography}

\begin{table*}
\centering
\caption{Orion A $\bf{B_{\text{LOS}}}$ values. Point numbers are as mapped in Fig.~\ref{OrionBottomAMap}. Negative values indicate magnetic fields pointed away from the observer and positive values are towards the observer. $\delta$Bs indicate the upper and lower limit uncertainties.}
\begin{tabular}{ c|c|c|c|c|c|c|c|c } 
 \hline
 \hline
 \makecell{Point \\Number} & RA ($^{\circ}$J2000) & Dec ($^{\circ}$J2000) & \Av\ (mag) & RM$_{\text{ON}}$ (rad~m$^{-2}$)& RM$_{\text{MC}}$ (rad~m$^{-2}$) & B ($\mu$G) & +$\delta$B ($\mu$G)& -$\delta$B ($\mu$G)\\ [0.5ex] 
 \hline\hline
1	&	80.56	&	-10.96	&	0.78	&	0.0	&	-1.4	&	-52	&	883	&	883	\\
2	&	81.08	&	-10.20	&	1.62	&	-25.7	&	-27.1	&	-284	&	315	&	323	\\
3	&	81.18	&	-10.15	&	1.30	&	-21.3	&	-22.7	&	-330	&	307	&	307	\\
4	&	81.25	&	-7.66	&	0.99	&	-41.6	&	-43.0	&	-991	&	620	&	620	\\
5	&	81.29	&	-7.13	&	1.30	&	-9.6	&	-11.0	&	-159	&	413	&	413	\\
6	&	81.65	&	-6.57	&	0.86	&	-33.1	&	-34.5	&	-1035	&	537	&	620	\\
7	&	81.89	&	-4.62	&	0.85	&	76.0	&	74.6	&	2327	&	896	&	965	\\
8	&	81.97	&	-11.14	&	0.94	&	29.4	&	28.0	&	700	&	517	&	463	\\
9	&	82.48	&	-4.89	&	0.90	&	8.5	&	7.1	&	196	&	558	&	558	\\
10	&	82.49	&	-10.98	&	0.85	&	-19.6	&	-21.0	&	-647	&	874	&	889	\\
11	&	82.76	&	-9.55	&	1.64	&	7.4	&	6.0	&	62	&	248	&	223	\\
12	&	82.89	&	-7.18	&	0.83	&	-36.3	&	-37.7	&	-1233	&	657	&	612	\\
13	&	83.81	&	-5.39	&	19.56	&	-13.5	&	-14.9	&	-23	&	38	&	38	\\
14	&	83.82	&	-5.38	&	21.47	&	10.9	&	9.5	&	15	&	36	&	36	\\
15	&	85.03	&	-5.22	&	2.09	&	81.0	&	79.6	&	595	&	215	&	215	\\
16	&	85.13	&	-10.38	&	1.81	&	28.1	&	26.7	&	242	&	265	&	265	\\
17	&	85.41	&	-5.70	&	1.62	&	58.9	&	57.5	&	607	&	168	&	168	\\
18	&	85.84	&	-5.04	&	1.58	&	45.9	&	44.5	&	483	&	275	&	275	\\
19	&	85.98	&	-10.51	&	2.43	&	26.3	&	24.9	&	154	&	121	&	121	\\
20	&	86.28	&	-5.49	&	3.11	&	24.1	&	22.7	&	104	&	114	&	114	\\
21	&	86.29	&	-6.58	&	1.29	&	-0.3	&	-1.7	&	-25	&	301	&	301	\\
22	&	86.31	&	-5.49	&	2.84	&	23.5	&	22.1	&	113	&	119	&	119	\\
23	&	86.32	&	-10.86	&	1.96	&	-3.1	&	-4.5	&	-37	&	220	&	220	\\
24	&	86.86	&	-4.30	&	2.05	&	13.5	&	12.1	&	93	&	167	&	167	\\
25	&	86.86	&	-4.30	&	2.05	&	10.6	&	9.2	&	71	&	157	&	157	\\
26	&	86.92	&	-11.49	&	0.77	&	-16.3	&	-17.7	&	-676	&	867	&	841	\\
27	&	87.12	&	-10.56	&	1.31	&	8.2	&	6.8	&	98	&	295	&	295	\\
28	&	87.18	&	-4.74	&	1.15	&	-19.6	&	-21.0	&	-370	&	456	&	455	\\
29	&	87.18	&	-8.05	&	1.66	&	42.6	&	41.2	&	418	&	308	&	308	\\
30	&	87.26	&	-5.04	&	0.85	&	-17.4	&	-18.8	&	-578	&	927	&	926	\\
31	&	87.61	&	-6.44	&	0.91	&	95.6	&	94.2	&	2553	&	695	&	698	\\
32	&	87.88	&	-7.94	&	0.77	&	103.6	&	102.2	&	3928	&	1135	&	1675	\\
33	&	87.88	&	-10.39	&	2.32	&	13.3	&	11.9	&	78	&	163	&	163	\\
34	&	87.89	&	-10.38	&	2.32	&	28.5	&	27.1	&	177	&	102	&	102	\\
\end{tabular}
\label{OrionABValues}
\end{table*}

\begin{table*}
\centering
\caption{Orion A $\bf{B_{\text{LOS}}}$ values, considering only the points that do not change direction within the estimated uncertainties. Point numbers are as mapped in Fig.~\ref{OrionBottomAMap}. Negative values indicate magnetic fields pointed away from the observer and positive values are towards the observer. $\delta$Bs indicate the upper and lower limit uncertainties.}
\begin{tabular}{ c|c|c|c|c|c|c|c|c } 
 \hline
 \hline
 \makecell{Point \\Number} & RA ($^{\circ}$J2000) & Dec ($^{\circ}$J2000) & \Av\ (mag) & RM$_{\text{ON}}$ (rad~m$^{-2}$)& RM$_{\text{MC}}$ (rad~m$^{-2}$) & B ($\mu$G) & +$\delta$B ($\mu$G)& -$\delta$B ($\mu$G)\\ [0.5ex] 
 \hline\hline
3	&	81.18	&	-10.15	&	1.30	&	-21.3	&	-22.7	&	-330	&	307	&	307	\\
4	&	81.25	&	-7.66	&	0.99	&	-41.6	&	-43.0	&	-991	&	620	&	620	\\
6	&	81.65	&	-6.57	&	0.86	&	-33.1	&	-34.5	&	-1035	&	537	&	620	\\
7	&	81.89	&	-4.62	&	0.85	&	76.0	&	74.6	&	2327	&	896	&	965	\\
8	&	81.97	&	-11.14	&	0.94	&	29.4	&	28.0	&	700	&	517	&	463	\\
12	&	82.89	&	-7.18	&	0.83	&	-36.3	&	-37.7	&	-1233	&	657	&	612	\\
15	&	85.03	&	-5.22	&	2.09	&	81.0	&	79.6	&	595	&	215	&	215	\\
17	&	85.41	&	-5.70	&	1.62	&	58.9	&	57.5	&	607	&	168	&	168	\\
18	&	85.84	&	-5.04	&	1.58	&	45.9	&	44.5	&	483	&	275	&	275	\\
19	&	85.98	&	-10.51	&	2.43	&	26.3	&	24.9	&	154	&	121	&	121	\\
29	&	87.18	&	-8.05	&	1.66	&	42.6	&	41.2	&	418	&	308	&	308	\\
31	&	87.61	&	-6.44	&	0.91	&	95.6	&	94.2	&	2553	&	695	&	698	\\
32	&	87.88	&	-7.94	&	0.77	&	103.6	&	102.2	&	3928	&	1135	&	1675	\\
34	&	87.89	&	-10.38	&	2.32	&	28.5	&	27.1	&	177	&	102	&	102	\\ 
\end{tabular}
\label{OrionABValuesModif}
\end{table*}

\begin{table*}
\centering
\caption{Orion B $\bf{B_{\text{LOS}}}$ values. Point numbers are as mapped in Fig.~\ref{OrionTopBMap}. Negative values indicate magnetic fields pointed away from the observer and positive values are towards the observer. $\delta$Bs indicate the upper and lower limit uncertainties.}
\begin{tabular}{ c|c|c|c|c|c|c|c|c }
 \hline
 \hline
 \makecell{Point \\Number} & RA ($^{\circ}$J2000) & Dec ($^{\circ}$J2000) & \Av\ (mag) & RM$_{\text{ON}}$ (rad~m$^{-2}$)& RM$_{\text{MC}}$ (rad~m$^{-2}$) & B ($\mu$G) & +$\delta$B ($\mu$G)& -$\delta$B ($\mu$G)\\ [0.5ex] 
 \hline\hline
 1	&	85.44	&	-1.92	&	37.36	&	-44.1	&	-76.4	&	-119	&	25	&	25	\\
2	&	85.45	&	-1.91	&	37.36	&	-50.1	&	-82.4	&	-129	&	28	&	28	\\
3	&	85.91	&	-2.95	&	2.03	&	82.6	&	50.3	&	393	&	199	&	199	\\
4	&	87.09	&	-1.29	&	2.84	&	56.0	&	23.7	&	122	&	125	&	125	\\
5	&	88.21	&	3.22	&	1.50	&	91.1	&	58.8	&	699	&	268	&	268	\\
6	&	88.92	&	-1.02	&	1.50	&	152.4	&	120.1	&	1423	&	268	&	263	\\
7	&	90.09	&	0.05	&	1.19	&	33.9	&	1.6	&	27	&	437	&	437	\\
8	&	90.15	&	-2.74	&	0.82	&	53.8	&	21.5	&	736	&	642	&	670	\\
\end{tabular}
\label{OrionBBValues}
\end{table*}

\begin{table*}
\centering
\caption{California $\bf{B_{\text{LOS}}}$ values. Point numbers are as mapped in Fig.~\ref{CaliforniaBMap}. Negative values indicate magnetic fields pointed away from the observer and positive values are towards the observer. $\delta$Bs indicate the upper and lower limit uncertainties.}
\begin{tabular}{ c|c|c|c|c|c|c|c|c } 
 \hline
 \hline
 \makecell{Point \\Number} & RA ($^{\circ}$J2000) & Dec ($^{\circ}$J2000) & \Av\ (mag) & RM$_{\text{ON}}$ (rad~m$^{-2}$)& RM$_{\text{MC}}$ (rad~m$^{-2}$) & B ($\mu$G) & +$\delta$B ($\mu$G)& -$\delta$B ($\mu$G)\\ [0.5ex] 
 \hline\hline
1	&	58.30	&	38.45	&	2.52	&	46.6	&	42.6	&	278	&	141	&	141	\\
2	&	58.63	&	41.91	&	0.90	&	-56.5	&	-60.5	&	-2106	&	1070	&	1070	\\
3	&	58.92	&	41.90	&	1.10	&	-7.5	&	-11.5	&	-255	&	539	&	539	\\
4	&	59.31	&	37.96	&	2.66	&	44.5	&	40.5	&	249	&	147	&	147	\\
5	&	59.60	&	41.87	&	1.81	&	-20.2	&	-24.2	&	-241	&	213	&	213	\\
6	&	59.75	&	40.05	&	1.02	&	24.9	&	20.9	&	541	&	615	&	615	\\
7	&	59.76	&	40.03	&	0.94	&	26.5	&	22.5	&	697	&	881	&	881	\\
8	&	59.87	&	41.78	&	2.45	&	12.9	&	8.9	&	60	&	228	&	228	\\
9	&	60.03	&	40.20	&	1.24	&	28.1	&	24.1	&	431	&	404	&	404	\\
10	&	60.19	&	37.80	&	1.16	&	73.0	&	69.0	&	1373	&	355	&	355	\\
11	&	60.54	&	41.51	&	2.88	&	-16.5	&	-20.5	&	-116	&	130	&	131	\\
12	&	60.92	&	38.33	&	2.06	&	7.1	&	3.1	&	26	&	255	&	255	\\
13	&	61.89	&	39.42	&	0.98	&	-3.4	&	-7.4	&	-210	&	701	&	701	\\
14	&	62.03	&	41.28	&	1.16	&	-2.4	&	-6.4	&	-129	&	637	&	637	\\
15	&	62.03	&	40.38	&	2.35	&	21.8	&	17.8	&	126	&	144	&	144	\\
16	&	62.14	&	41.56	&	0.98	&	-2.4	&	-6.4	&	-180	&	669	&	669	\\
17	&	62.24	&	36.59	&	0.93	&	59.1	&	55.1	&	1761	&	827	&	826	\\
18	&	62.34	&	38.81	&	2.07	&	56.0	&	52.0	&	430	&	138	&	138	\\
19	&	63.69	&	36.04	&	0.69	&	55.4	&	51.4	&	4160	&	3259	&	3438	\\
20	&	64.47	&	35.28	&	1.16	&	34.8	&	30.8	&	614	&	394	&	394	\\
21	&	64.57	&	38.01	&	3.55	&	-24.2	&	-28.2	&	-131	&	119	&	119	\\
22	&	64.62	&	38.04	&	3.05	&	-13.7	&	-17.7	&	-94	&	175	&	175	\\
23	&	65.06	&	38.83	&	1.37	&	27.3	&	23.3	&	348	&	384	&	384	\\
24	&	65.33	&	35.19	&	0.78	&	31.5	&	27.5	&	1402	&	1799	&	1799	\\
25	&	65.39	&	38.61	&	1.73	&	5.1	&	1.1	&	11	&	259	&	259	\\
26	&	65.61	&	39.93	&	1.08	&	-17.4	&	-21.4	&	-495	&	656	&	656	\\
27	&	65.97	&	34.86	&	0.78	&	41.0	&	37.0	&	1925	&	1257	&	1257	\\
28	&	66.09	&	39.35	&	1.76	&	-26.4	&	-30.4	&	-313	&	189	&	190	\\
29	&	66.10	&	39.37	&	1.76	&	-18.0	&	-22.0	&	-226	&	286	&	286	\\
30	&	66.77	&	36.06	&	2.32	&	-21.2	&	-25.2	&	-181	&	214	&	214	\\
31	&	66.92	&	38.39	&	1.47	&	-3.6	&	-7.6	&	-102	&	360	&	360	\\
32	&	67.98	&	34.94	&	1.69	&	-44.4	&	-48.4	&	-529	&	240	&	240	\\
33	&	68.28	&	35.84	&	2.17	&	-24.2	&	-28.2	&	-220	&	171	&	171	\\
34	&	68.77	&	36.78	&	2.12	&	-14.0	&	-18.0	&	-145	&	227	&	227	\\
35	&	68.87	&	35.25	&	1.63	&	-17.6	&	-21.6	&	-248	&	268	&	268	\\ 
\end{tabular}
\label{CaliforniaBValues}
\end{table*}

\begin{table*}
\centering
\caption{California $\bf{B_{\text{LOS}}}$ values, considering only the points that do not change direction within the estimated uncertainties. Point numbers are as mapped in Fig.~\ref{CaliforniaBMap}. Negative values indicate magnetic fields pointed away from the observer and positive values are towards the observer. $\delta$Bs indicate the upper and lower limit uncertainties.}
\begin{tabular}{ c|c|c|c|c|c|c|c|c }
 \hline
 \hline
 \makecell{Point \\Number} & RA ($^{\circ}$J2000) & Dec ($^{\circ}$J2000) & \Av\ (mag) & RM$_{\text{ON}}$ (rad~m$^{-2}$)& RM$_{\text{MC}}$ (rad~m$^{-2}$) & B ($\mu$G) & +$\delta$B ($\mu$G)& -$\delta$B ($\mu$G)\\ [0.5ex] 
 \hline\hline
1	&	58.30	&	38.45	&	2.52	&	46.6	&	42.6	&	278	&	141	&	141	\\
2	&	58.63	&	41.91	&	0.90	&	-56.5	&	-60.5	&	-2106	&	1070	&	1070	\\
4	&	59.31	&	37.96	&	2.66	&	44.5	&	40.5	&	249	&	147	&	147	\\
5	&	59.60	&	41.87	&	1.81	&	-20.2	&	-24.2	&	-241	&	213	&	213	\\
9	&	60.03	&	40.20	&	1.24	&	28.1	&	24.1	&	431	&	404	&	404	\\
10	&	60.19	&	37.80	&	1.16	&	73.0	&	69.0	&	1373	&	355	&	355	\\
17	&	62.24	&	36.59	&	0.93	&	59.1	&	55.1	&	1761	&	827	&	826	\\
18	&	62.34	&	38.81	&	2.07	&	56.0	&	52.0	&	430	&	138	&	138	\\
19	&	63.69	&	36.04	&	0.69	&	55.4	&	51.4	&	4160	&	3259	&	3438	\\
20	&	64.47	&	35.28	&	1.16	&	34.8	&	30.8	&	614	&	394	&	394	\\
21	&	64.57	&	38.01	&	3.55	&	-24.2	&	-28.2	&	-131	&	119	&	119	\\
27	&	65.97	&	34.86	&	0.78	&	41.0	&	37.0	&	1925	&	1257	&	1257	\\
28	&	66.09	&	39.35	&	1.76	&	-26.4	&	-30.4	&	-313	&	189	&	190	\\
32	&	67.98	&	34.94	&	1.69	&	-44.4	&	-48.4	&	-529	&	240	&	240	\\
33	&	68.28	&	35.84	&	2.17	&	-24.2	&	-28.2	&	-220	&	171	&	171	\\
\end{tabular}
\label{CaliforniaBValuesModif}
\end{table*}

\begin{table*}
\centering
\caption{Perseus $\bf{B_{\text{LOS}}}$ values. Point numbers are as mapped in Fig.~\ref{PerseusBMap}. Negative values indicate magnetic fields pointed away from the observer and positive values are towards the observer. $\delta$Bs indicate the upper and lower limit uncertainties.}
\begin{tabular}{ c|c|c|c|c|c|c|c|c }
 \hline
 \hline
 \makecell{Point \\Number} & RA ($^{\circ}$J2000) & Dec ($^{\circ}$J2000) & \Av\ (mag) & RM$_{\text{ON}}$ (rad~m$^{-2}$)& RM$_{\text{MC}}$ (rad~m$^{-2}$) & B ($\mu$G) & +$\delta$B ($\mu$G)& -$\delta$B ($\mu$G)\\ [0.5ex] 
 \hline\hline
1	&	50.01	&	29.69	&	2.62	&	59.2	&	28.1	&	194	&	176	&	176	\\
2	&	50.01	&	31.13	&	1.27	&	89.6	&	58.5	&	1229	&	457	&	624	\\
3	&	50.15	&	30.72	&	3.50	&	40.1	&	9.0	&	47	&	128	&	116	\\
4	&	51.29	&	31.48	&	2.34	&	35.1	&	4.0	&	32	&	101	&	101	\\
5	&	51.80	&	29.02	&	1.75	&	26.2	&	-4.9	&	-57	&	268	&	269	\\
6	&	51.91	&	31.40	&	2.87	&	15.1	&	-16.0	&	-100	&	157	&	158	\\
7	&	52.03	&	29.44	&	1.16	&	18.9	&	-12.2	&	-310	&	530	&	537	\\
8	&	52.15	&	29.37	&	2.32	&	24.8	&	-6.3	&	-50	&	103	&	106	\\
9	&	52.17	&	30.83	&	4.21	&	60.2	&	29.1	&	134	&	67	&	67	\\
10	&	52.54	&	30.55	&	3.75	&	-8.3	&	-39.4	&	-196	&	109	&	109	\\
11	&	52.54	&	28.65	&	1.37	&	6.5	&	-24.6	&	-441	&	339	&	338	\\
12	&	52.61	&	30.06	&	0.64	&	28.1	&	-3.0	&	-238	&	1351	&	1363	\\
13	&	52.92	&	28.68	&	1.89	&	19.9	&	-11.2	&	-118	&	246	&	253	\\
14	&	53.17	&	31.86	&	1.50	&	67.4	&	36.3	&	557	&	488	&	359	\\
15	&	53.57	&	31.20	&	4.09	&	28.2	&	-2.9	&	-13	&	88	&	88	\\
16	&	53.88	&	30.54	&	2.11	&	22.2	&	-8.9	&	-80	&	146	&	147	\\
17	&	54.13	&	32.31	&	1.03	&	34.3	&	3.2	&	113	&	435	&	443	\\
18	&	54.41	&	32.13	&	0.92	&	175.4	&	144.3	&	7160	&	1121	&	4585	\\
19	&	54.46	&	31.25	&	2.42	&	110.8	&	79.7	&	605	&	207	&	211	\\
20	&	54.47	&	30.93	&	3.08	&	48.7	&	17.6	&	103	&	83	&	81	\\
21	&	54.91	&	32.91	&	2.03	&	126.8	&	95.7	&	904	&	291	&	256	\\
22	&	55.04	&	32.15	&	2.69	&	200.0	&	168.9	&	1137	&	466	&	103	\\
23	&	55.52	&	30.34	&	1.25	&	19.2	&	-11.9	&	-257	&	473	&	475	\\
24	&	55.81	&	31.25	&	3.47	&	39.9	&	8.8	&	46	&	94	&	94	\\
\end{tabular}
\label{PerseusBValues}
\end{table*}

\begin{table*}
\centering
\caption{Perseus $\bf{B_{\text{LOS}}}$ values, considering only the points that do not change direction within the estimated uncertainties. Point numbers are as mapped in Fig.~\ref{PerseusBMap}. Negative values indicate magnetic fields pointed away from the observer and positive values are towards the observer. $\delta$Bs indicate the upper and lower limit uncertainties.}
\begin{tabular}{ c|c|c|c|c|c|c|c|c }
 \hline
 \hline
 \makecell{Point \\Number} & RA ($^{\circ}$J2000) & Dec ($^{\circ}$J2000) & \Av\ (mag) & RM$_{\text{ON}}$ (rad~m$^{-2}$)& RM$_{\text{MC}}$ (rad~m$^{-2}$) & B ($\mu$G) & +$\delta$B ($\mu$G)& -$\delta$B ($\mu$G)\\ [0.5ex] 
 \hline\hline
1	&	50.01	&	29.69	&	2.62	&	59.2	&	28.1	&	194	&	176	&	176	\\
2	&	50.01	&	31.13	&	1.27	&	89.6	&	58.5	&	1229	&	457	&	624	\\
9	&	52.17	&	30.83	&	4.21	&	60.2	&	29.1	&	134	&	67	&	67	\\
10	&	52.54	&	30.55	&	3.75	&	-8.3	&	-39.4	&	-196	&	109	&	109	\\
11	&	52.54	&	28.65	&	1.37	&	6.5	&	-24.6	&	-441	&	339	&	338	\\
14	&	53.17	&	31.86	&	1.50	&	67.4	&	36.3	&	557	&	488	&	359	\\
18	&	54.41	&	32.13	&	0.92	&	175.4	&	144.3	&	7160	&	1121	&	4585	\\
19	&	54.46	&	31.25	&	2.42	&	110.8	&	79.7	&	605	&	207	&	211	\\
20	&	54.47	&	30.93	&	3.08	&	48.7	&	17.6	&	103	&	83	&	81	\\
21	&	54.91	&	32.91	&	2.03	&	126.8	&	95.7	&	904	&	291	&	256	\\
22	&	55.04	&	32.15	&	2.69	&	200.0	&	168.9	&	1137	&	466	&	103	\\
\end{tabular}
\label{PerseusBValuesModif}
\end{table*}

\end{document}